\documentclass[conference]{IEEEtran}
\IEEEoverridecommandlockouts
\usepackage{cite}
\usepackage{amsmath,amssymb,amsfonts}

\usepackage{pifont} 
\usepackage{xcolor}
\usepackage{xspace}
\usepackage{algorithmic}
\usepackage{graphicx}
\usepackage{textcomp}
\usepackage{booktabs}   
\usepackage{threeparttable}
\usepackage{multirow}
\usepackage{subfigure}
\usepackage{tabularx}
\usepackage[ruled,vlined,linesnumbered,noend]{algorithm2e}
\usepackage{enumitem}
\usepackage{tikz}
\usepackage[hyphens]{url}
\usepackage{hyperref}
\usepackage{comment}
\usepackage{adjustbox}
\usepackage{soul}

\def\BibTeX{{\rm B\kern-.05em{\sc i\kern-.025em b}\kern-.08em
    T\kern-.1667em\lower.7ex\hbox{E}\kern-.125emX}}
\begin{document}

\pdfpagewidth=8.5in
\pdfpageheight=11in

\newcommand{\iscasubmissionnumber}{NaN}

\newcommand\name{XtraMAC\xspace}
\newcommand{\yessymbol}{\textcolor{green!70!black}{\ding{51}}} 
\newcommand{\nosymbol}{\textcolor{red}{\ding{55}}}              

\newcommand{\yesnox}{
  \textcolor{green!70!black}{\ding{51}}%
  \kern-0.7em
  \raisebox{0.3ex}{\rotatebox[origin=c]{55}{\textcolor{black}{\textbf{/}}}}
}


\pagenumbering{arabic}

\title{\name: An Efficient MAC Architecture for Mixed-Precision LLM Inference on FPGA}



\author{
\IEEEauthorblockN{
    Feng Yu\textsuperscript{*},
    Hongshi Tan\textsuperscript{*},
    Yao Chen\textsuperscript{\dag,\ddag},
    Weng-Fai Wong\textsuperscript{*},
    Bingsheng He\textsuperscript{*,\ddag}
}
\IEEEauthorblockA{
    \textsuperscript{*}\textit{School of Computing, National University of Singapore}, Singapore \\
    \textsuperscript{\dag}\textit{School of Computer Science and Technology, Huazhong University of Science and Technology}, Wuhan, China \\
    \{yuf, hongshi\}@u.nus.edu, \{dcswwf, dcsheb\}@nus.edu.sg, chenyao\_cs@hust.edu.cn
}
\thanks{\textsuperscript{\ddag}Yao~Chen and Bingsheng~He are the corresponding authors.}
}

\maketitle
\thispagestyle{plain}
\pagestyle{plain}

\begin{abstract} 
The widespread adoption of mixed-precision quantization in large language models (LLMs) has created demand for hardware that can efficiently perform multiply–accumulate (MAC) operations across mixed datatypes and switch datatypes at runtime. Existing FPGA-based MAC solutions fall short due to limitations in fixed-datatype design, inefficient spatial or temporal resource sharing, and poor support for mixed-precision execution. These limitations collectively lead to under-utilization of DSP resources, limiting achievable parallelism and throughput. 
In this work, we present \name{}, a novel MAC architecture that unifies integer, floating-point, and mixed-precision operations within a single, datatype-adaptive microarchitecture. \name{} decomposes all supported MAC formats into a shared integer mantissa product with lightweight sign and exponent handling, enabling dynamic operand packing and efficient DSP resource sharing with constant latency and initiation interval of one across all datatypes.
Evaluated on an AMD Xilinx U55c FPGA, \name{} achieves 1.4–2.0$\times$ higher compute density, reduces per-operation LUT, FF, and DSP consumption by 27–51\%, and delivers up to 1.9$\times$ greater energy efficiency and 1.2$\times$ speedup on representative mixed-precision LLM workloads. The implementation of \name{} is open-sourced at \url{https://github.com/Xtra-Computing/XtraMAC}.
\end{abstract}


\section{Introduction}

Multiply–accumulate (MAC) operations are the fundamental building blocks for a broad spectrum of computational tasks, ranging from digital signal processing to deep learning. In recent years, large language models (LLMs) have emerged as the most prominent and demanding consumers of MAC computation, as their inference workloads are dominated by matrix multiplications involving billions of parameters. This explosive growth in MAC demand has catalyzed the adoption of low-precision quantization techniques to reduce both memory footprint and computational energy. Quantization methods~\cite{frantar2022gptq, lin2024awq, dettmers2024spqr, smoothquant, MLSYS2024_atom, chee2023quip} target different model components (weights, activations, MoE experts) with mixed numeric formats and assign different datatypes across layers based on their quantization sensitivity, e.g., lower-bit integers for compute-intensive layers and floating-point formats for numerically sensitive ones~\cite{frantar2022gptq, lin2024awq, smoothquant}. As summarized in Table~\ref{tab:patterns}, this results in diverse MAC datatype combinations across different quantization schemes and model components.

\begin{figure}[t]
\centering
\setlength{\abovecaptionskip}{0.2cm}
\setlength{\belowcaptionskip}{-0.2cm}
\includegraphics[width=0.48\textwidth]{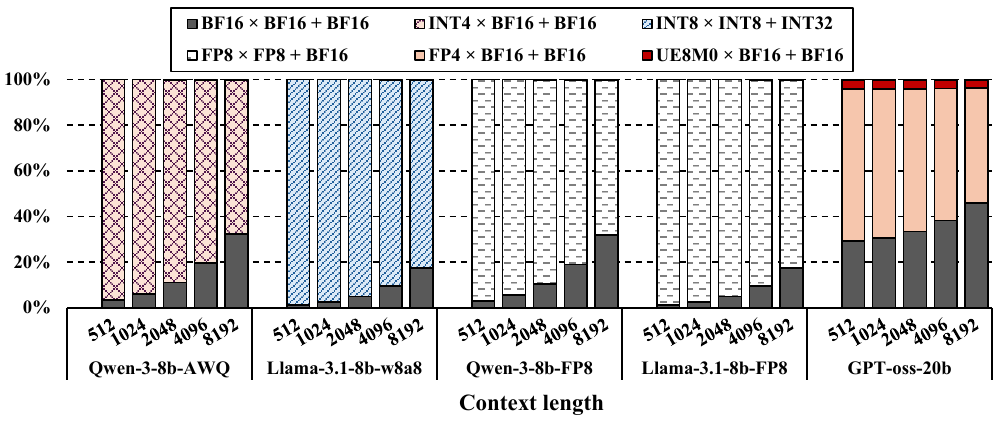}
\caption{Distribution of MAC operations during the decode stage for various quantized LLM checkpoints in Table~\ref{tab:quantized_usecases_downloads} across different context lengths. Each segment represents a unique MAC configuration; non-MAC operations are omitted as they account for less than 1\% of total operations.}
\label{fig:mac-breakdown}
\end{figure}

However, this proliferation of mixed-precision quantization introduces a new challenge for hardware design. We define a MAC operation as P = A$\times$B + C, and identify two distinct computational patterns that arise in practice. The first is \textbf{mixed-precision MAC}, in which the multiplicands $A$ and $B$ are represented in heterogeneous numeric formats or bitwidths (e.g., INT4$\times$BF16). The second is \textbf{runtime datatype switching}, in which a single hardware unit must alternate among distinct MAC datatypes as execution traverses different model components. For example, a single forward pass may transition from INT4$\times$BF16 in projection layers to BF16$\times$BF16 in attention layers. As shown in Figure~\ref{fig:mac-breakdown}, Qwen-3-8B-AWQ executes over 68\% of its decode-stage MACs in INT4$\times$BF16 for projection layers while its attention layers retain BF16$\times$BF16, exemplifying the coexistence of both patterns within a single model. This dynamic heterogeneity demands hardware that can natively support both patterns without performance degradation.

\begin{table}[t]
\centering
\caption{Diversity of MACs across LLM quantization schemes.}
\vspace{-0.3cm}
\label{tab:patterns}
\resizebox{\linewidth}{!}{
\begin{threeparttable}
\begin{tabular}{c|c|c|c}
\toprule
\textbf{Category} 
& \textbf{Examples} 
& \textbf{Projection / FFN MACs} 
& \textbf{Attention MACs} \\
\midrule
Weight-only quant. 
& AWQ, GPTQ, SpQR 
& $\text{INT}\times\text{FP}+\text{FP}\rightarrow\text{FP}$ 
& $\text{FP}\times\text{FP}+\text{FP}\rightarrow\text{FP}$ \\
\midrule
Weight–act quant. 
& SmoothQuant, Atom 
& $\text{INT}\times\text{INT}+\text{INT}\rightarrow\text{INT}$ 
& $\text{FP}\times\text{FP}+\text{FP}\rightarrow\text{FP}$ \\
\midrule
Native LLMs 
& GPT-oss-20b, 120b 
& $\text{MXFP4/BF16}\tnote{*}\times\text{FP}+\text{FP}\rightarrow\text{FP}$
& $\text{FP}\times\text{FP}+\text{FP}\rightarrow\text{FP}$ \\
\bottomrule
\end{tabular}

\begin{tablenotes}[flushleft]\small
\item[*] In GPT-oss models, the MoE blocks use MXFP4, others use BF16 datatypes.
\end{tablenotes}

\end{threeparttable}
}
\end{table}

\begin{table*}[t]
\centering
\resizebox{\linewidth}{!}{
\begin{threeparttable}
\caption{Survey of FPGA-based MAC designs for mixed precision (MP) and runtime datatype switching (RDS) support.}
\vspace{-0.4cm}
\label{tab:fpga-mac}
\begin{tabular}{lcccccccl}
\toprule
\textbf{Name} & \textbf{Year} & \textbf{DataType(A)} & \textbf{DataType(B)} & \textbf{DataType(C)} & \textbf{DataType(P)} & \textbf{MP} & \textbf{RDS} & \textbf{Note} \\
\midrule
FINN~\cite{umuroglu2017finn} & 2017 & Binary & Binary & Integer & Integer & \nosymbol & \nosymbol & Fixed architecture for binary neural networks \\
FP-BNN~\cite{liang2017fpbnn} & 2017 & Binary & Binary & Integer & Integer & \nosymbol & \nosymbol & Fixed architecture for binary neural networks \\
DNNBuilder~\cite{zhang2018dnnbuilder} & 2018 & Fixed-point & Fixed-point & Fixed-point & Fixed-point & \nosymbol & \nosymbol & Precision set at synthesis time \\
BISMO~\cite{bismo2018}\tnote{*} & 2019 & Integer & Integer & Integer & Integer & \yessymbol & \yessymbol & Runtime switchable integer bit width \\
Xilinx FP Operator~\cite{floating_point_ip}\tnote{\dag} & 2020 & \textbf{Floating-point} & \textbf{Floating-point} & \textbf{Floating-point} & \textbf{Floating-point} & \nosymbol & \nosymbol & Format chosen at synthesis time \\
Triple MAC~\cite{kerner2021triple} & 2021 & Fixed-point & Fixed-point & Fixed-point & Fixed-point & \nosymbol & \nosymbol & Static precision; width chosen at design time \\
TATAA~\cite{tataa2023}\tnote{\ddag} & 2025 & INT8 or BF16 & INT8 or BF16 & INT32 or BF16 & INT32 or BF16 & \nosymbol & \yessymbol & Runtime switchable between INT8 and BF16 \\
\hline
\multirow{2}{*}{\textbf{Ours}} & \multirow{2}{*}{2025} & \textbf{Integer,} &  \textbf{Integer,} & \textbf{Integer,} & \textbf{Integer,} & \multirow{2}{*}{\yessymbol} & \multirow{2}{*}{\yessymbol} & \multirow{2}{*}{Runtime switchable across all datatypes} \\
 &  & \textbf{floating-point} & \textbf{floating-point} & \textbf{floating-point} & \textbf{floating-point} & & & \\
\bottomrule
\end{tabular}
\begin{tablenotes}[para,flushleft]
\item[*] BISMO supports arbitrary integer bit-width selection at runtime, but the maximum permissible precision is fixed at synthesis time due to FPGA resource constraints.
\item[\dag] Operands \(A,B,C\) and output \(P\) must share the same floating point datatype.  
\item[\ddag] Operands \(A,B,C\) and output \(P\) must share the same datatype, either all INT8 or all BF16.  
\end{tablenotes}
\end{threeparttable}
}
\end{table*}

Field-programmable gate arrays (FPGAs) have emerged as a promising platform for addressing these challenges, owing to their ability to implement customized compute pipelines with fine-grained, bit-level control over arithmetic and dataflow. In particular, the datapaths leading to the primary computational units in FPGAs, the DSP cores, can be tailored to accommodate diverse datatype requirements.
Nevertheless, existing solutions for mixed-precision MAC support on FPGAs remain suboptimal. Conventional approaches to leveraging DSPs in FPGAs for mixed-precision computation and runtime datatype switching can be broadly categorized as follows:
\begin{itemize} [leftmargin=*]
    \item \textbf{Operand upcasting}, which promotes low-precision operands to match a fixed high-precision MAC unit, resulting in significant waste of the DSP bit space;
    \item \textbf{Spatial replication or temporal sharing}, which instantiates multiple datatype-specific datapaths or reuses a single datapath across cycles to support runtime datatype switching, leading to low effective DSP utilization, as only a subset of resources is active at any moment.
\end{itemize}

For example, the AMD Xilinx Floating-Point Operator~\cite{floating_point_ip} achieves average \textbf{32.4\%} DSP bit-utilization efficiency when executing low-precision workloads, as all operands must be upcast into a fixed high-precision floating-point format. For runtime datatype switching, the spatial-replication based design duplicates multiple datatype-specific datapaths using the same Floating-Point Operator IP~\cite{floating_point_ip} further reduces efficiency: only one datapath is active at a time while the others remain idle, resulting in an effective DSP utilization of average \textbf{26.7\%}.
Temporal-sharing architectures such as TATAA~\cite{tataa2023} decompose a BF16 MAC into four sequential INT8 operations, yielding 71.1\% utilization for INT8-based MAC computation but only \textbf{8.9\%} effective utilization for BF16-based MAC computation due to the multi-cycle decomposition.


The root cause of these inefficiencies is a theoretical disconnect between the processing patterns and the fixed resource granularity of FPGA DSP slices. While DSP packing has been shown to substantially boost utilization for integer-only workloads~\cite{hikonv, hipack}, current solutions lack the bit-level analysis needed to extend this technique to current workloads. As a result, each DSP delivers throughput far below its arithmetic ceiling, and this shortfall widens as LLMs increasingly mix precisions and switch datatypes within a single forward pass.

In this work, we bridge this gap by introducing a resource-compact MAC architecture that is natively aware of mixed precision and supports runtime datatype switching within a unified datapath. Our key contributions are:
\begin{itemize}[leftmargin=*]
    \item We present a unified formulation showing that the multiplication component of integer, floating-point, and mixed-precision MACs can all be decomposed into an integer mantissa product with lightweight sign and exponent handling.
    \item We propose \name{}, a datatype-adaptive MAC architecture that decouples format interpretation from arithmetic execution, unifying integer, floating-point, and mixed-precision MAC operations within a single shared datapath.
    \item We develop a DSP-centric design principle that exploits dynamic bit mapping and multi-lane packing to maximize multiplier utilization, and adopt a fixed four-stage pipeline to sustain a constant latency and initiation interval of one across all supported datatypes.
    \item We demonstrate that \name improves compute density by 1.4–2.0× and reduces LUT, FF, DSP usage by 30.0\%, 47.9\%, 50.0\% compared with state-of-the-art FPGA baselines, achieving 1.2× lower GEMV latency and 1.9× higher energy efficiency compared with GPU baseline.
\end{itemize}

\section{Background and Design Motivation}

\label{sec:motivation}

Motivated by the mixed-precision and runtime datatype-switching demands of LLM workloads, we first formalize DSP utilization under widely adopted conventions, and then review prevailing MAC microarchitectures in FPGA-based solutions, with particular emphasis on their DSP utilization.

\begin{figure*}[htp]
  \centering
  \includegraphics[width=0.98\textwidth]{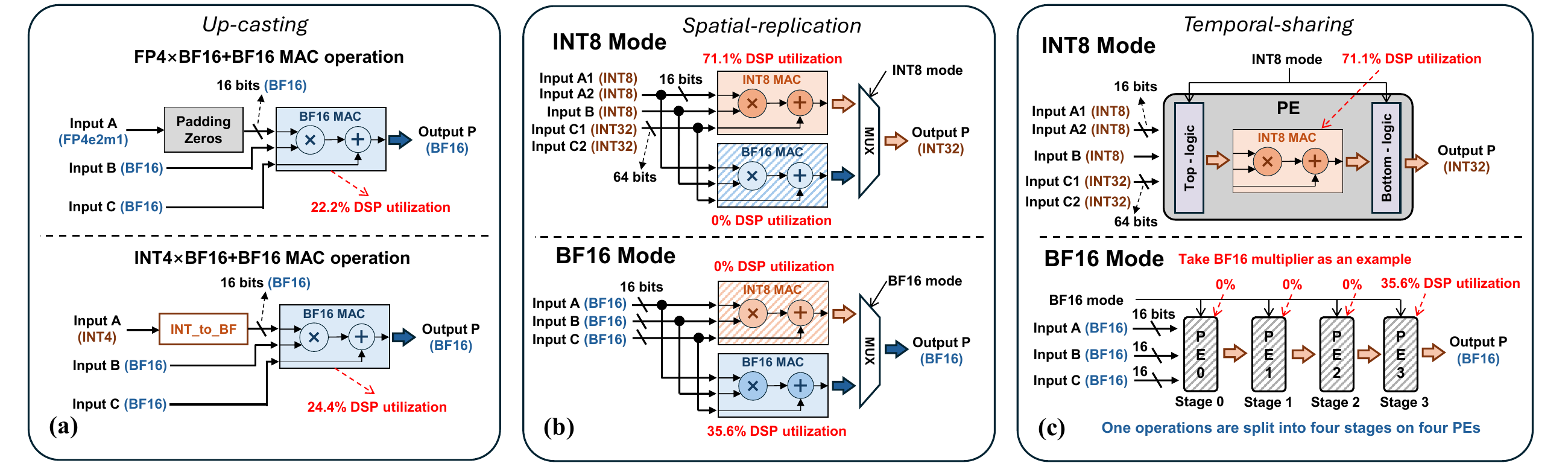}
  \vspace*{-0.2cm}
  \caption{Overview of existing FPGA-based MAC architectures supporting mixed precision and runtime datatype switching. (a) Upcasting-based method using FP Operator~\cite{floating_point_ip} for mixed precision. (b) Spatial replication for multi-datatype support~\cite{floating_point_ip}. (c) Temporal-sharing based multi-datatype support (TATAA~\cite{tataa2023}).}
  \label{fig:compare}
\end{figure*}

\subsection{DSP Utilization}

In modern FPGAs, hardened DSP slices center on a dedicated multiplier that serves as the primary arithmetic resource and performs integer multiplication. The associated pre-addition, post-addition, and pipeline logic is structurally simpler and incurs substantially lower area and power overheads compared with the multiplier itself\cite{xilinx_dsp48e2_ug958, ug579}. Therefore, following the operand-bit-based utilization model in
\cite{chen2023m4bram}, we quantify DSP utilization based on how effectively the multiplier hardware is exercised. Let $w_a$ and $w_b$ denote the effective bit-widths of the multiplicands
involved in an operation, we define the DSP utilization as
\[
U_{\mathrm{DSP}} = (w_a + w_b) / W_{\mathrm{mul}},
\]
where $W_{\mathrm{mul}}$ is the sum of the two input-port widths of the DSP multiplier. In this paper, we target the DSP48E2 primitive widely deployed in modern Xilinx FPGAs, whose multiplier accepts a 27-bit A-port operand and an 18-bit B-port operand, giving $W_{\mathrm{mul}} = 45$ bits~\cite{xilinx_dsp48e2_ug958}.

\subsection{Current MAC designs in FPGA-based solutions}

MAC designs in FPGA-based solutions have evolved to support a variety of numerical formats, ranging from binary, integer, fixed-precision arithmetic to full-precision floating-point.
As summarized in Table \ref{tab:fpga-mac}, these designs demonstrate the flexibility of reconfigurable logic for arithmetic specialization but reveal persistent limitations in mixed-precision support and runtime datatype switching.

\subsubsection{Lack of Efficient Mixed-Precision Support}

\begin{figure}[t]
\centering
\setlength{\abovecaptionskip}{0.2cm}
\setlength{\belowcaptionskip}{-0.2cm}
\includegraphics[width=0.46\textwidth]{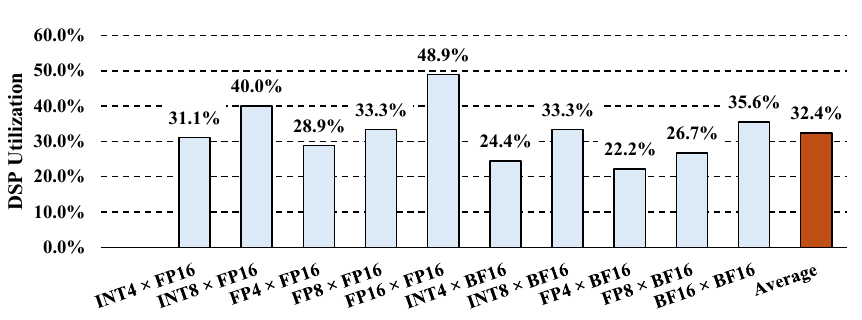}
\caption{DSP utilization of the upcasting-based design under different mixed-precision datatype combinations (FP8 = E4M3, FP4 = E2M1).}
\label{fig:dsp_mixed}
\end{figure}

Early FPGA-based accelerators, such as FINN~\cite{umuroglu2017finn}, FP-BNN~\cite{liang2017fpbnn}, and DNNBuilder~\cite{zhang2018dnnbuilder}, target fixed integer or fixed-point formats with precision determined at synthesis time. Consequently, most FPGA MAC architectures are optimized for fixed or uniform operand types and provide limited support for mixed-precision computation. In practice, mixed-precision operations are typically handled by upcasting low-precision operands to the highest supported precision and executing them on a high-precision MAC datapath. As illustrated in Fig.~\ref{fig:compare}(a), and exemplified by the AMD Xilinx Floating-Point Operator~\cite{floating_point_ip}, low-precision operands are padded or promoted to match the highest supported precision, and all operations are executed on a fixed high-precision datapath. This design results in significant hardware inefficiency: a large portion of the DSP bit capacity remains unused, as quantified in Fig.~\ref{fig:dsp_mixed}.




\begin{figure}[t]
\centering
\setlength{\abovecaptionskip}{0.2cm}
\setlength{\belowcaptionskip}{-0.2cm}
\includegraphics[width=0.46\textwidth]{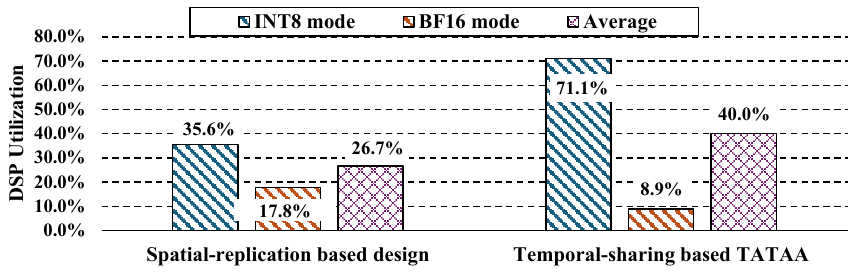}
\caption{DSP utilization comparison on existing FPGA-based MAC architectures supporting runtime datatype switching in Fig.\ref{fig:compare}.}
\label{fig:dsp_runtime}
\end{figure}

\subsubsection{Inefficient DSP Utilization and Limitations in Runtime Datatype Switching}

Recent designs supporting runtime datatype switching on FPGAs predominantly adopt one of two microarchitectural strategies: spatial replication or temporal sharing. Representative designs are illustrated in Fig.~\ref{fig:compare}(b) and (c). While each approach offers certain advantages, both introduce inherent limitations and inefficiencies in DSP utilization, particularly when accommodating the diverse numeric requirements of LLM workloads.

\textbf{Spatial Replication.} As shown in Fig.~\ref{fig:compare}(b), designs such as those described in~\cite{arora2022tensor} address runtime datatype switching by duplicating MAC units for each datatype and employing a multiplexer (MUX) to select the active datapath at runtime. For example, an INT8/BF16-configurable instance instantiates both INT8 and BF16 MACs, toggling between them with a control signal. While this method enables zero-latency switching and avoids pipeline bubbles, it incurs substantial hardware overhead as the number of supported formats increases. Idle MAC units for inactive datatype remain unused, leading to poor resource efficiency and failing to leverage the concurrency benefits inherent to low-precision computation. In Fig.~\ref{fig:dsp_runtime}, the average DSP utilization drops to 26.7\%.

\textbf{Temporal Sharing.} As shown in Fig.~\ref{fig:compare}(c), a more sophisticated approach in TATAA~\cite{tataa2023}, which reuses integer MAC units for both INT8 and BF16 computation by decomposing each BF16 operation into a sequence of INT8 micro-operations executed over multiple cycles. This strategy avoids the need for separate BF16 logic, reducing area overhead. However, it comes at the expense of spatial parallelism and throughput: each BF16 operation monopolizes four processing elements (PEs) and pipeline stages, effectively capping peak BF16 throughput at one quarter of that achievable for INT8, given the same hardware footprint.
As quantified in Fig.~\ref{fig:dsp_runtime}, the DSP utilization drops to 8.9\% when supporting BF16-based MAC, which significantly reduces the hardware effectiveness.

These pronounced inefficiencies in DSP utilization underscore a fundamental microarchitectural problem, one that becomes increasingly problematic as mixed-precision floating-point computation and runtime datatype switching grow prevalent in state-of-the-art LLM inference workloads.




\subsection{Our Observations}


The above quantitative results point to a common root cause: a mismatch between the bit-level processing patterns of low-precision MAC operations and the fixed resource granularity of FPGA DSP slices. Existing designs either upcast low-bit operands to a wide format or serialize high-precision operations over a narrow low-precision core. In both cases, 
\textbf{the DSP multiplier is used as an opaque}, single-lane primitive, rather than as a bit-space that can be systematically partitioned and shared across datatypes.
Furthermore, 
both approaches \textbf{treat datatype switching as a coarse control problem over whole datapaths}, instead of a fine-grained organization of workload allocation within the DSP bit-space.

These observations suggest that substantial further gains are unlikely to be achieved by incremental datapath tweaks alone. Instead, we rethink mixed-precision MAC support from a processing-pattern perspective: how low- and mixed-precision INT/FP multiplications decompose into bit-level operations that can be allocated to the DSP multiplier based on its given processing characteristics, and how the DSP utilization could be improved with proper parallelism supported by the microarchitecture of the MAC. The next section develops this processing-pattern formulation, which forms the theoretical basis for the \name architecture and its ability to achieve high DSP utilization and fine-grained runtime datatype switch.

\section{Processing Pattern Formulation}
\label{sec:design_principles}
\setlength{\abovedisplayskip}{4pt}
\setlength{\belowdisplayskip}{4pt}
\setlength{\abovedisplayshortskip}{3pt}
\setlength{\belowdisplayshortskip}{3pt}

To fully exploit the sharing and performance potential of mixed-precision MAC architectures, we first analyze how low- and mixed-precision MAC operations map onto the native multipliers and accumulators of DSP slices in FPGAs. Unlike conventional fixed-precision designs, these MAC operations introduce heterogeneous operand widths that misalign with DSP bit-granularity, giving rise to distinct sub-DSP packing and lane parallelism patterns. Hence, a resource-compact MAC architecture must be grounded in a systematic characterization of these bit-level processing patterns. 
We begin with normalized, non-exceptional values to establish the core formulation, and extend the analysis in a later subsection to cover special values and rounding. 
Under these conventions, XtraMAC produces bit-exact results matching NVIDIA A100/H100 Tensor Cores~\cite{nvidia_a100_ds, nvidia2022h100} and the official AMD Floating-Point Operator~\cite{floating_point_ip} across all supported datatypes.

\subsection{Shared Multiplication Datapath across Different Datatypes}
\label{subsec:principles_multiplier}

For a floating-point operand $x = s_x \cdot 2^{e_x} \cdot m_x$ in IEEE format, we assume that the mantissa $m_x$ includes the implicit leading-one bit (i.e., $m_x \in [1,2)$ for normalized values). The product of two such values $x$ and $y$ naturally decomposes as
\begin{equation}
    x \cdot y
    = (s_x \oplus s_y)
      \cdot 2^{\,e_x + e_y - \text{bias}}
      \cdot (m_x \cdot m_y),
    \label{eq:fp_mul}
\end{equation}
when adopting the MAC constructed with FPGA DSP slice, the DSP computes only the mantissa product $m_x m_y$, while sign and exponent are processed separately~\cite{xilinx_dsp48e2_ug958, floating_point_ip}. After multiplication, the product mantissa is normalized by a leading-zero count (LZC):
\begin{equation}
    \Delta = \mathrm{LZC}(m_x m_y), \quad
    m^{\text{norm}} = (m_x m_y) \ll \Delta,
\end{equation}
\begin{equation}
    e^{\text{out}} = e_x + e_y - \text{bias} - \Delta.
\end{equation}

For mixed-precision INT$\times$FP computation, the integer operand $a$ is interpreted as a two's complement value and decomposed into sign and magnitude:
\[
    a = s_a \cdot m_a,
\]
where $m_a$ is treated as an integer mantissa after sign extraction. Since an integer carries no exponent encoding, we assign it a logical unbiased exponent of zero, which corresponds to a biased exponent value equal to the bias of the floating-point output format (e.g., 127 for FP32 and BF16, or 15 for FP16). The floating-point operand $y$ is represented same as the $x$, with $m_y$ including the leading-one bit. Their product becomes
\begin{equation}
    a \cdot y
    = (s_a \oplus s_y)
      \cdot 2^{\,e_y - \mathrm{bias}}
      \cdot (m_a \cdot m_y),
    \label{eq:intfp_mul}
\end{equation}
so that the DSP multiplier again computes only the mantissa product $m_a m_y$. The exponent path forwards $e_y$, subtracts the format bias, and applies the normalization shift:
\begin{equation}
    \Delta = \mathrm{LZC}(m_a m_y), \quad
    m^{\text{norm}} = (m_a m_y) \ll \Delta,
\end{equation}
\begin{equation}
    e^{\text{out}} = e_y - \mathrm{bias} - \Delta.
\end{equation}

Eqs.~\eqref{eq:fp_mul} and~\eqref{eq:intfp_mul} show that both FP$\times$FP and INT$\times$FP multiplication share the same core operation: the DSP computes an integer mantissa product, while sign and exponent are handled outside the DSP. This indicates that a common multiplication datapath can be shared across different datatypes. The only differences arise in how operands are mapped into \((s,m,e)\) fields and how the exponent is updated after normalization. Hence, across all integer, floating-point, and mixed-precision formats, multiplication can be described as follows:
\begin{enumerate}[leftmargin=*]
\item \textbf{Mapping:} extract or construct sign, mantissa, and exponent for each input operand of the DSP slice.
\item \textbf{DSP multiplication:} compute product \(m_{\text{prod}} = m_A \cdot m_B\).
\item \textbf{Post-compute:} apply LZC-based normalization and update the exponent according to the operand formats.
\end{enumerate}

This formulation shows that datatype-specific behavior is confined to lightweight mapping and post-compute logic, while the DSP slice consistently performs the same integer multiplication. Consequently, a single multiplier datapath can support FP$\times$FP, INT$\times$FP, and pure integer multiplication with minimal additional hardware.


\subsection{Datatype-specific Accumulation}
\label{subsec:principles_adders}

Unlike the multiplication stage, which naturally shares a unified integer--mantissa structure across all data types, addition behaves fundamentally differently due to incompatible computation patterns arising from their distinct bit semantics. For integer addition, two's-complement integers $x, y$ satisfy
\[
    x + y = (x \oplus y) \oplus \mathrm{carry}(x,y),
\]
which maps directly to the FPGA's ripple-carry chain. A $w_{\text{int}}$-bit integer adder therefore exhibits linear resource cost:
\begin{equation}
    C_{\mathrm{int}}(w_{\text{int}}) \approx \alpha_{\mathrm{int}} \, w_{\text{int}},
    \label{eq:int_cost}
\end{equation}
where $\alpha_{\mathrm{int}}$ denotes the LUT cost per bit imposed by the carry-chain fabric.

For floating-point addition, given $x = s_x 2^{e_x} m_x$ and $y = s_y 2^{e_y} m_y$, define the exponent gap
\[
    \Delta e = e_x - e_y.
\]
Assuming $e_x \ge e_y$, the smaller mantissa must be right-shifted by an \emph{arbitrary} distance determined at runtime:
\[
    m_s = m_x \pm \bigl(m_y \cdot 2^{-\Delta e}\bigr).
\]
The resulting mantissa must then be normalized using LZC:
\[
    \Delta = \mathrm{LZC}(m_s),\qquad
    m_{\mathrm{out}} = (m_s \ll \Delta),\qquad
    e_{\mathrm{out}} = e_x - \Delta.
\]

Because $\Delta e$ may take any value in the exponent range, the alignment step must support a variable-distance right shift across the full mantissa width $w_{\text{fp}}$; similarly, normalization requires a variable-distance left shift. Implementing such shifts on FPGA requires a logarithmic barrel shifter. A classical barrel shifter for an $w_{\text{fp}}$-bit mantissa contains $\log_2 w_{\text{fp}}$ stages of $w_{\text{fp}}$ multiplexers~\cite{schulte2005barrel}, giving
\[
    N_{\mathrm{MUX}} = w_{\text{fp}} \log_2 w_{\text{fp}},
\]
and the LUT cost scales superlinearly:
\begin{equation}
    C_{\mathrm{shifter}}(w_{\text{fp}}) \approx \beta_{\mathrm{sh}} \, w_{\text{fp}} \log_2 w_{\text{fp}},
    \label{eq:shifter_cost}
\end{equation}
where $\beta_{\mathrm{sh}}$ denotes the LUT cost per multiplexer stage. Prior work confirms that these alignment and normalization shifters dominate the LUT footprint of FPGA floating-point adders~\cite{moctar2012reducing,ehliar2014area}.

Thus, these characteristics highlight the inherently datatype-specific nature of accumulation. Integer addition typically requires a wide bit width $w_{\text{int}}$ (for example, 32 bits for INT4/INT8 accumulation~\cite{nvidia_ampere_whitepaper, nvidia2022h100}) but avoids expensive shifting by relying on efficient carry-chain logic. Floating-point addition operates on narrower mantissas $w_{\text{fp}}$ (for example, 10-bit for FP16, 7-bit for BF16) but incurs substantial cost due to alignment and normalization shifters. 
Consequently, a unified INT–FP adder would force integer additions to traverse the same alignment and normalization shifters as floating-point additions, even though integer addition fundamentally requires no shifting. This wastes shifter area, which must be sized for the wider integer width $w_{\text{int}}$, making it significantly less resource-efficient than maintaining separate adder paths.

\subsection{Parallel Mixed-Precision MACs on a Single DSP Slice}

In FPGA implementation, a DSP slice computes a single wide integer multiplication
\[
    P_{\text{DSP}} = A_{\text{DSP}} \cdot B_{\text{DSP}},
\]
where $A_{\text{DSP}}$ and $B_{\text{DSP}}$ are bit vectors on the two input ports. Conventional fixed-precision designs place each operand in the least-significant bits, leaving most of the multiplier unused and resulting in low DSP bit utilization. However, as shown in Section~\ref{subsec:principles_multiplier}, all MAC datatypes ultimately reduce to integer mantissa multiplication after appropriate operand mapping. This enables multiple low-precision lanes to be packed into disjoint bit regions of the DSP inputs, thereby exploiting the full multiplier width~\cite{hikonv, hipack, xilinx_int8_wp486}.

To perform this packing, the mapping stage assigns each mantissa or integer magnitude to a non-overlapping bit range:
\begin{equation}
    A_{\text{DSP}} = \sum_i (a_i \ll s_i), \qquad
    B_{\text{DSP}} = \sum_j (b_j \ll t_j),
    \label{eq:dsp_pack}
\end{equation}
where $s_i$ and $t_j$ are per-lane shift offsets selected to avoid cross-lane interference. These offsets depend on the datatype combination (INT, FP, or INT$\times$FP), but the DSP always receives well-formed integer operands regardless of precision.

Once packed, the DSP performs one wide multiplication, and all lane products appear at predetermined bit positions:
\begin{equation}
    P_{\text{DSP}}
    = \sum_{i,j} (a_i b_j) \ll (s_i + t_j).
    \label{eq:dsp_cross}
\end{equation}
In the post-compute stage, each lane product is extracted using a fixed shift-and-mask operation:
\begin{equation}
    P_{i,j}
    = \left( P_{\text{DSP}} \gg (s_i + t_j) \right)
      \,\&\, (2^{S}-1),
    \label{eq:lane_recover}
\end{equation}
where $S$ is the bit stride allocated per lane. Let $W_{\text{lane}}$ denote the maximum bit-width of any supported product $|a_i \cdot b_j|$. To guarantee no overlap, the stride must satisfy
\[
    S \ge W_{\text{lane}} + G,
\]
where $G$ is a small guard margin (typically one bit) used to absorb carries.

By incorporating lane packing into the mapping stage and integrating product extraction into the post-compute stage, the unified multiplication datapath of Section~\ref{subsec:principles_multiplier} naturally extends to support parallel mixed-precision MACs. As a result, multiple INT, FP, or INT$\times$FP operations can be executed in parallel within a single DSP slice by sharing the same datapath across different datatypes.

\begin{figure*}[t]
\centering
\vspace{-0.2cm}
\includegraphics[width=\textwidth]{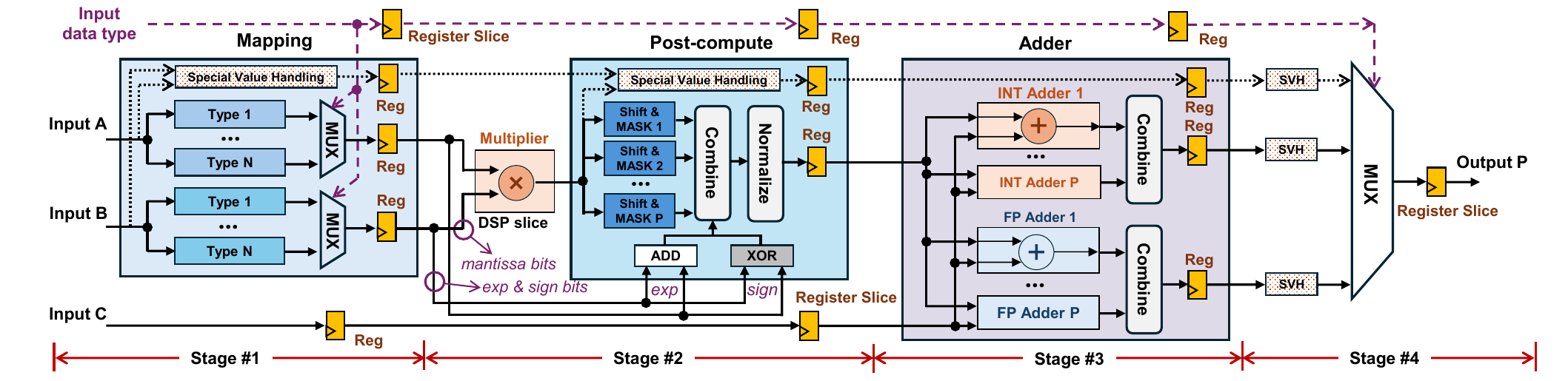}
\caption{Overview of the \name{} architecture supporting $N$ datatype combinations and up to $P$-way parallelism. (SVH: special value handling).}
\label{fig:arch_overview}
\end{figure*}

The parallelism of shared DSP multiplier is constrained by DSP input widths. For a given datatype, once its per-lane stride $S$ is determined, the maximum achievable parallelism is
\begin{equation}
    \mathrm{Parallelism} \le \min\!\left(
        \left\lfloor \frac{L_A}{S} \right\rfloor,\;
        \left\lfloor \frac{L_B}{S} \right\rfloor
    \right),
    \label{eq:parallelism}
\end{equation}
where $L_A$ = 27 and $L_B$ = 18 are the DSP48E2 input widths.
Importantly, different datatypes (e.g., INT4$\times$BF16, FP8$\times$FP8, INT8$\times$INT8) have different operand precisions and thus correspond to different stride values $S$. Once the stride for a datatype is fixed, the achievable parallelism for that datatype follows directly from Eq.~\eqref{eq:parallelism} and depends solely on $L_A$, $L_B$.

In summary, by reducing all numerical formats to products and packing these lanes into the DSP inputs at bit-precise offsets, \name{} enables a single DSP multiplier to be shared across integer, floating-point, and mixed-precision MACs. The polynomial structure of packed multiplication in Eq.~\eqref{eq:dsp_cross} preserves strict lane isolation, while per-lane normalization and exponent reconstruction restore correct floating point and mixed-precision semantics. The bit-level control forms the basis of our mixed-precision support and allows multiple low-precision MAC lanes to execute in parallel.

\subsection{Special Value Handling}

\subsubsection{Numerical Coverage}
XtraMAC adopts flush-to-zero (FTZ) and denormals-are-zero (DAZ) semantics throughout the computation datapath, consistent with the numerical conventions widely employed in modern floating-point compute hardware, such as NVIDIA A100/H100 Tensor Cores~\cite{nvidia_a100_ds, nvidia2022h100} and the AMD Xilinx Floating-Point Operator~\cite{floating_point_ip}. Subnormal inputs are treated as zero upon ingestion and outputs falling below the minimum value are flushed to zero. NaN inputs propagate as canonical quiet NaN (qNaN), infinity is preserved with its sign, and conflicting cases such as $\infty \times 0$ and $+\infty + (-\infty)$ resolve to qNaN. For formats that do not encode infinity, all-ones exponent encodings are treated as NaN. Integer-to-floating-point conversion is exact, and floating-point accumulation applies round-to-nearest-even (RN-even) throughout.

\subsubsection{Pipeline-Invariant Exception Handling}
A key design requirement is that exception handling must not disrupt the timing behavior of the main datapath. XtraMAC achieves this by detecting all special cases, including NaN, infinity, subnormal, and overflow, at the input and encoding them as status flags that are forwarded through the same matched register slices as the operands, ensuring that control and data remain temporally aligned throughout the pipeline. At the output, the appropriate result is selected through purely combinational logic, requiring no stalls, pipeline flushes, or control flow divergence. Overflow is resolved by saturating the result to $\pm\infty$ through the same flag selection mechanism, ensuring uniform treatment across all exception types. 
Thus, the pipeline's latency and throughput are preserved unconditionally, independent of whether inputs are normal or exceptional.

\section{\name Architecture Details}\label{sec:architecture}

Guided by the processing-pattern formulation in Section~\ref{sec:design_principles}, \name separates numeric interpretation from numeric execution: lightweight peripheral logic performs format decoding, bit-mapping, packing, and post-computation, while the arithmetic core comprises a datatype-invariant DSP multiplier and decoupled INT/FP accumulation paths. The architecture is parameterized by (i) the number of supported datatypes $N$, chosen at synthesis time, and (ii) the maximum parallelism $P$ across these $N$ datatypes, which defines the uniform per-lane structure shared by all stages and is chosen no larger than the bound in Eq.~\ref{eq:parallelism}. This section details how \name realizes a fully pipelined mixed-precision MAC datapath with runtime datatype switching.

\subsection{Overall Architecture}

Figure~\ref{fig:arch_overview} illustrates the four-stage pipeline of \name. 
A dedicated datatype-select signal is registered at pipeline entry and propagated through all four stages via matched delay slices, selecting one of the $N$ supported datatypes each cycle.
All datatype-specific mapping and reconstruction submodules are instantiated statically, and runtime switching is achieved entirely through input datatype controlled multiplexing without any form of reconfiguration. The pipeline processes up to $P$ logical lanes per operation, where $P$ corresponds to the maximum parallelism among the $N$ supported datatypes, ensuring that all formats share a common parallel substrate. The four stages perform operand interpretation and DSP packing (Stage~1), datatype-invariant multiplication and per-lane post-computation (Stage~2), datatype-specific accumulation via decoupled INT/FP adder paths (Stage~3), and final output selection and assembly (Stage~4). The following subsections describe each stage in detail.

\subsection{Stage~1: Operand Interpretation and Bit-Mapping}

Stage~1 receives input operands $A$ and $B$, which may encode multiple concatenated low-precision values, and translates them into datatype-specific bit-packed DSP-port operands for Stage~2. The datatype input identifies the active format among the $N$ supported datatypes and determines how each data value is interpreted. To support $N$ datatypes without reconfiguration, Stage~1 instantiates $N$ parallel mapping submodules. Each submodule decodes $A$ and $B$ according to its datatype definition. Floating-point submodules extract the sign, exponent, and mantissa fields, restore the implicit leading~1, and pack the mantissa bits into the designated DSP input positions, while forwarding the exponent and sign fields as metadata. Integer submodules perform two’s-complement decoding, pack the magnitude bits into the DSP inputs, and forward the sign bit. For pure integer datatypes, the exponent metadata is set to zero, effectively treating the value as a fixed-point mantissa with exponent $0$. After all submodules compute in parallel, the datatype signal selects a single pair of bit-packed DSP-port operands and the corresponding per-lane sign and exponent metadata, providing a uniform interface to the multiplication module in Stage~2 and enabling cycle-level switching across heterogeneous datatypes.

\subsection{Stage~2: Multiplier and Post-Compute}

Stage~2 receives the bit-packed DSP-port operands and the associated per-lane metadata produced by Stage~1. The DSP-based multiply module is fully datatype-invariant: the DSP slice performs a pure integer multiplication, consistent with the decomposition in Eq.~\ref{eq:fp_mul} and Eq.~\ref{eq:intfp_mul}, which shows that all supported INT/FP formats reduce to a mantissa-level product with separate exponent and sign handling. The resulting product from the DSP slice aggregates all packed lanes into a single wide bitfield. The post-compute module reconstructs the $P$ logical lanes by applying deterministic per-lane shift-and-mask extraction based on the packing layout generated in Stage~1. Floating-point lanes perform leading-zero counting, normalization shifts, and exponent updates, while integer lanes require only magnitude extraction and sign XOR without normalization. The post-compute module instantiates $P$ identical per-lane pipelines so that all lanes are reconstructed in parallel, and Stage~2 outputs $P$ integer or floating-point values for accumulation in Stage~3.

\subsection{Stage~3: Datatype-specific Accumulation}

Stage~3 receives the per-lane products from Stage~2 together with the corresponding per-lane accumulator operands $C$ and performs accumulation using decoupled adder modules. The integer module implements a bank of two’s-complement adders, while the floating-point module performs exponent alignment, mantissa add/subtract, and renormalization. This structural separation, following the principles in Section~\ref{sec:design_principles}, avoids the area and latency overheads of a unified INT--FP adder. Both modules operate every cycle and compute $P$ lane results in parallel; the datatype signal then selects the output.

\subsection{Stage~4: Output Selection}

Stage~4 receives the lane-wise accumulated results and selects the output based on the input data type. The $P$ lane results are concatenated into the final packed output word (e.g., four FP8 lanes or two BF16 lanes forming 32 bits). No additional normalization or conversion is required, and the stage emits one multi-lane MAC result per cycle.

\subsection{Pipeline Behavior}

\name adopts a fixed four-stage logical pipeline in which each stage contains a bounded combinational block followed by a register boundary, cleanly separating logic evaluation from temporal sequencing. The DSP slice is configured with its internal pipeline registers disabled so that the multiplication behaves as a purely combinational block between Stage~1 and Stage~2 registers. All interface signals are time-aligned through matched delay slices to sustain continuous throughput. For example, the input datatype signal, consumed in both Stage~1 and Stage~4, is delayed by the appropriate number of register slices before reaching Stage~4, ensuring that control and data correspond to the same dynamic operation. Likewise, the input operand $C$, required only in Stage~3, is delayed to align with the products emerging from Stage~2.

XtraMAC fixes the pipeline at four logical stages but leaves the cycle count of each stage configurable at synthesis time. By default, every stage completes in a single clock cycle, yielding an end-to-end latency of four cycles. When a stage with complex logic (e.g., Stage 3) becomes the critical path, the designer can insert additional pipeline registers within it to trade latency for a higher clock frequency. This requires only extending the matched delay slices on parallel paths; the initiation interval remains one. By combining per-stage-configurable latency with pipeline-aligned propagation, XtraMAC provides a deterministic latency and an initiation interval of one across all supported datatypes.


\subsection{System Integration}
Although \name{} adds an input data type port for runtime datatype selection, this signal functions solely as pipeline-aligned control metadata and does not modify the operand interface or timing behavior of a standard MAC unit. All datatype-dependent formatting and reconstruction logic resides inside the \name{} pipeline, so external modules continue to provide and consume the same lane-packed operands used in existing accelerators. With a fixed latency and an initiation interval of one across all datatypes, \name{} can replace a conventional MAC unit without requiring any schedule or interface changes, enabling easy system integration.

\section{Evaluation}
In this section, we evaluate \name{} in terms of its mixed-precision coverage and runtime datatype adaptability. We further compare its resource efficiency and performance against state-of-the-art FPGA baselines, including the AMD Xilinx Floating-Point Operator~\cite{floating_point_ip} and TATAA~\cite{tataa2023}. 

\subsection{Experimental Setting}

All designs, including \name{} and the baselines, are implemented using Vivado~2022.2 and Vitis~2022.2 on the AMD Xilinx Alveo U55c FPGA card. To ensure a fair comparison, we apply the same tool configurations and compilation settings across all implementations. For floating-point operations, we adopt the round-to-nearest-even (RN-even) rounding mode and flush subnormal inputs to zero. For integer operations, we use two's-complement signed representation, apply saturation on overflow, and truncate extra bits during accumulation.

\subsection{Mixed-Precision Evaluation}

\begin{figure*}[t]
\centering
\vspace{-0.2cm}
\includegraphics[width=\textwidth]{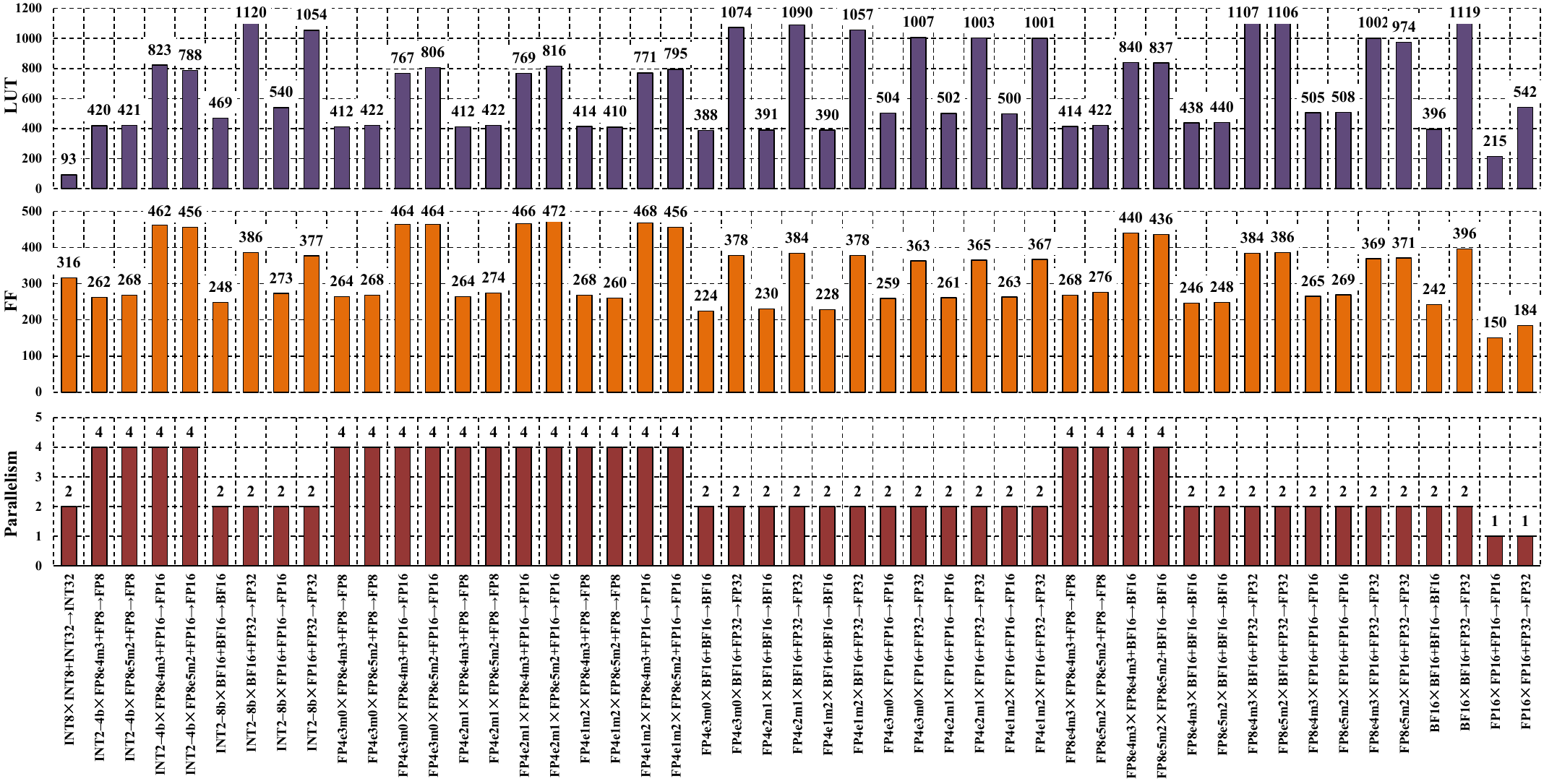}
\caption{Resource consumption and parallelism of \name across all $A \times B + C \rightarrow P$ configurations. For brevity, constant metrics (DSP = 1, latency = 4 cycles, II = 1) are omitted. The full exponent–mantissa formats (e.g., E4M3, E5M2) are shown only at the first occurrence of each data type configuration.}
\label{fig:mixed_precision_matrix}
\end{figure*}

We evaluate \name under various datatype configurations to analyze its mixed-precision capability and adaptive parallelism. In this subsection, each row in Fig.~\ref{fig:mixed_precision_matrix} corresponds to a standalone MAC instance synthesized with a single active datatype configuration, and \name's runtime datatype switching mechanism is disabled. Fig.~\ref{fig:mixed_precision_matrix} reports resource consumption for LUT, FF, and DSP usage, along with performance metrics including latency, initiation interval, and achievable parallelism. Across all configurations, \name maintains a four-cycle latency and an initiation interval of one, indicating that datatype variation does not affect pipeline depth. In general, lower-precision formats such as FP4 and FP8 achieve higher packing parallelism (up to four MAC lanes per DSP), whereas higher-precision formats like BF16 are limited to two lanes per DSP because their wider mantissas reduce spatial packing opportunities.

For detailed resource characterization, the integer mode (INT8$\times$INT8$\rightarrow$INT32) provides two-way parallelism with the lowest LUT usage among all evaluated configurations. Floating-point configurations require substantially more LUTs and FFs due to the mantissa-alignment logic in the floating-point adder, where the barrel shifter dominates resource consumption. As the adder bitwidth increases, both the shift range and the shifter complexity grow super-linearly, leading to a significant increase in LUT and FF utilization.

\subsection{Runtime Datatype Switching Evaluation}

\begin{table}[t]
\centering
\caption{Resource consumption for runtime-switching evaluation.}
\label{tab:exp-rps-settings}
\footnotesize
\setlength{\tabcolsep}{6pt}
\renewcommand{\arraystretch}{1.1}
\vspace{-0.2cm}
\resizebox{\linewidth}{!}{
\begin{threeparttable}
\begin{tabular}{c | c | c | c c c}
\toprule
\textbf{Config ID} & \textbf{Data Type Combinations} & \textbf{Model Name(s)} & \textbf{LUT} & \textbf{FF} & \textbf{DSP} \\
\midrule
\multirow{2}{*}{I} & INT4$\times$BF16 + BF16        & \multirow{2}{*}{Qwen3-8b-AWQ}               & \multirow{2}{*}{436} & \multirow{2}{*}{302} & \multirow{2}{*}{1} \\
                   & BF16$\times$BF16 + BF16        &                                              &                      &                      &                     \\[2pt]
\midrule
\multirow{2}{*}{II} & INT8$\times$INT8 + INT32       & \multirow{2}{*}{Llama-3.1-8b-W8A8}          & \multirow{2}{*}{568} & \multirow{2}{*}{513} & \multirow{2}{*}{1} \\
                   & BF16$\times$BF16 + BF16        &                                              &                      &                      &                     \\[2pt]
\midrule
\multirow{2}{*}{III} & FP8$\times$FP8 + BF16          & Qwen3-8b-FP8                                 & \multirow{2}{*}{948} & \multirow{2}{*}{622} & \multirow{2}{*}{1} \\
                   & BF16$\times$BF16 + BF16        & Llama-3.1-8b-FP8                             &                      &                      &                     \\[2pt]
\midrule
\multirow{2}{*}{IV} & FP4$\times$BF16 + BF16         & \multirow{2}{*}{GPT-oss-20b\tnote{$\dagger$}} & \multirow{2}{*}{395} & \multirow{2}{*}{274} & \multirow{2}{*}{1} \\
                   & BF16$\times$BF16 + BF16        &                                              &                      &                      &                     \\[2pt]
\bottomrule
\end{tabular}

\begin{tablenotes}[flushleft]\small
\item[$\dagger$] The multiplication of UE8M0 with BF16 in GPT-oss-20b is implemented by offsetting the BF16 exponent, so it is not treated as a separate MAC operation.
\end{tablenotes}
\end{threeparttable}
}
\end{table}

\begin{figure}[t]
  \centering
  \setlength{\abovecaptionskip}{0.2cm}
  \setlength{\belowcaptionskip}{-0.2cm}
  \includegraphics[width=0.48\textwidth]{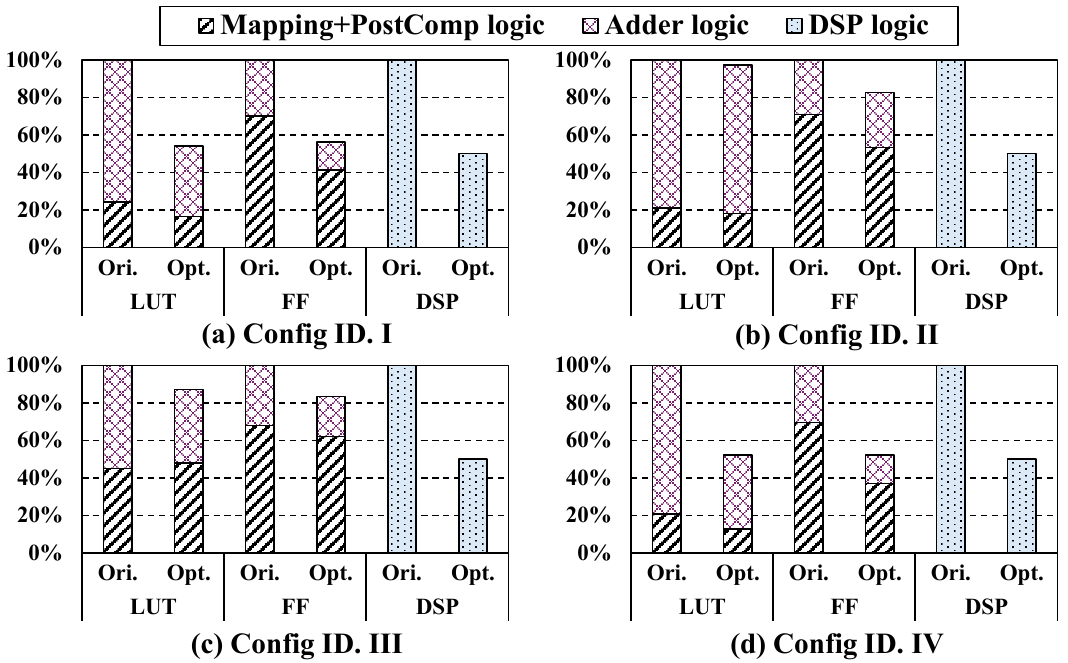}
  \caption{Normalized resource breakdown for different Config IDs in Table~\ref{tab:exp-rps-settings}.}
  \label{fig:rps_breakdown}
  \vspace{-0.2cm}
\end{figure}

We evaluate the hardware efficiency of runtime datatype switching using representative datatype combinations derived from Fig.~\ref{fig:mac-breakdown}. Table~\ref{tab:exp-rps-settings} summarizes the post-synthesis resource consumption of \name{} for all evaluated cases. Across these configurations, \name{} maintains a constant DSP cost, shares LUT and FF usage across the supported datatypes, and preserves a latency of four clock cycles and an initiation interval of one. To quantify the benefits of resource sharing, Fig.~\ref{fig:rps_breakdown} further breaks down LUT and FF usage into mapping and post-processing logic, arithmetic logic, and adder logic.

Config~I reuses both the arithmetic core and the adder unit, while the mapping logic remains separate because integer and floating-point operands require different alignment procedures. As a result, the LUT and FF usage of the adder logic is reduced by approximately 50\% relative to a naive design that instantiates separate MAC units for each datatype, whereas the mapping cost shows almost no change.  
Config~II shares only the arithmetic core, as INT32 and BF16 accumulations follow different normalization and saturation rules. 
Moreover, the mapping logic cannot be reused because integer formats do not include exponent alignment, leading to negligible resource reduction.  
Config~III enables partial reuse of the BF16 adder lanes because FP8 and BF16 operate at different levels of vector parallelism, four lanes versus two lanes. In this case, two of the four BF16 adder lanes can be shared directly to FP8 adders, which provides approximately 30\% LUT and FF savings in the adder logic.
Config~IV achieves near-complete reuse across the mapping, arithmetic, and adder units. Both MAC types operate with a parallelism of two, and FP4 operands can be expanded to BF16 by zero-padding without requiring exponent alignment or rounding adjustment. This compatibility enables full hardware sharing and results in the lowest LUT and FF utilization among all configurations.


\subsection{Scalability Evaluation of XtraMAC}

\begin{figure}[t]
  \centering
  \setlength{\abovecaptionskip}{0.2cm}
  \setlength{\belowcaptionskip}{-0.2cm}
  \includegraphics[width=0.48\textwidth]{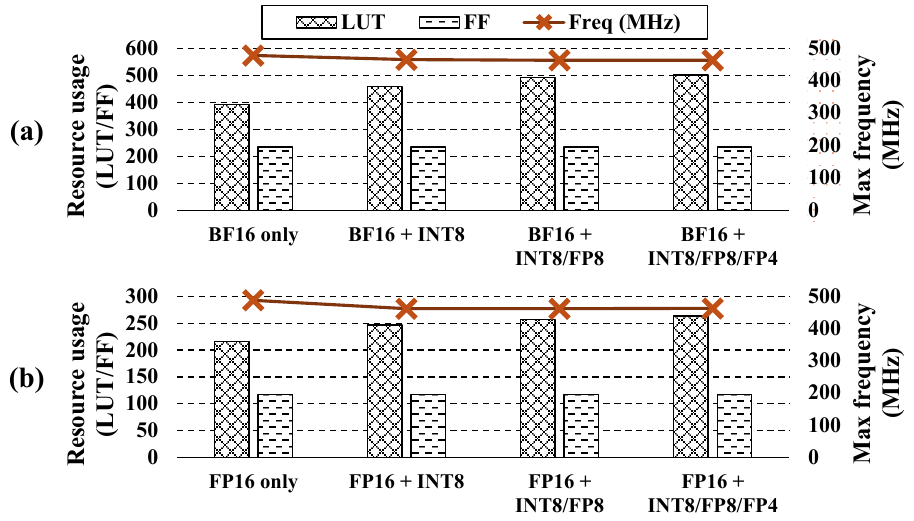}
  \caption{Scalability evaluation of XtraMAC with increasing mixed-precision datatype support, using a single DSP slice. (a) FP16-based MAC, progressively adding $\mathrm{INT8}$, $\mathrm{FP8}$, and $\mathrm{FP4}$ multiplier operand datatypes. (b) BF16-based MAC, progressively adding the same multiplier operand datatypes.}
  \label{fig:scalability}
  \vspace{-0.3cm}
\end{figure}

To evaluate the scalability of XtraMAC, we progressively increase the number of supported datatype combinations starting from BF16 and FP16 MAC baselines, incrementally enabling INT8, FP8 (E4M3), and FP4 (E2M1) mixed-precision modes. At each stage, we synthesize the design and measure FPGA resource utilization and maximum frequency. As shown in Fig.~\ref{fig:scalability}, LUT usage increases gradually as additional datatypes are enabled, due to the extra peripheral logic required for operand mapping and post-compute normalization. DSP usage remains constant across all configurations, confirming that the core multiplier is fully shared across all formats. The maximum frequency decreases slightly, from 483\,MHz to 462\,MHz, indicating that XtraMAC scales gracefully with the number of supported datatypes.

\subsection{Comparison with SOTA MAC IPs on FPGAs}

\subsubsection{Mixed-precision comparison}
We compare \name{} with the AMD Xilinx Floating-Point Operator IP~\cite{floating_point_ip}, the vendor’s standard MAC implementation for floating-point data formats. Both designs use a non-blocking AXI interface, RN-even rounding, and flush subnormal values to zero. All implementations operate with an initiation interval (II) of 1 and same latency for a fair comparison, and resource utilization is reported after synthesis. Since the vendor operator does not support mixed-precision computation, we augment it with a commonly used open-source integer-to-floating-point conversion module~\cite{hartman1996mmachine,int_to_float} to enable mixed-precision inputs.

\begin{table}[t]
\centering
\footnotesize
\caption{Normalized resource utilization comparison by parallelism.}
\label{tab:mixed_precision_comparison}
\vspace{-0.2cm}
\resizebox{\linewidth}{!}{
\begin{threeparttable}
\begin{tabular}{
>{\centering\arraybackslash}p{0.8cm}
>{\centering\arraybackslash}p{0.8cm}
>{\centering\arraybackslash}p{0.8cm}
>{\centering\arraybackslash}p{0.8cm}
|
>{\centering\arraybackslash}p{0.6cm}
|
>{\centering\arraybackslash}p{1.4cm}
>{\centering\arraybackslash}p{1.4cm}
|
>{\centering\arraybackslash}p{0.8cm}
>{\centering\arraybackslash}p{1.3cm}
}
\toprule
\textbf{Type(A)$^\dagger$} & \textbf{Type(B)} & \textbf{Type(C)} & \textbf{Type(P)} 
& \textbf{Res.} & \textbf{VendorIP~\cite{floating_point_ip}} 
& \textbf{\name{}$^\ddagger$} & \textbf{Red.(\%)} & \textbf{Comp.Den.(\(\uparrow\))} \\
\midrule
\multirow{3}{*}{INT2-8} & \multirow{3}{*}{BF16} & \multirow{3}{*}{BF16} & \multirow{3}{*}{BF16}
& LUT & 331 & 235 & 29.0\% & 1.4x \\
 &  &  &  & FF  & 222 & 124 & 44.1\% & 1.8x \\
 &  &  &  & DSP & 1   & 0.5 & 50.0\% & 2.0x \\
\midrule
\multirow{3}{*}{INT2-8} & \multirow{3}{*}{FP16} & \multirow{3}{*}{FP16} & \multirow{3}{*}{FP16}
& LUT & 387 & 270 & 30.2\% & 1.4x \\
 &  &  &  & FF  & 262 & 137 & 47.7\% & 1.9x \\
 &  &  &  & DSP & 1   & 0.5 & 50.0\% & 2.0x \\
\midrule
\multirow{3}{*}{FP4} & \multirow{3}{*}{BF16} & \multirow{3}{*}{BF16} & \multirow{3}{*}{BF16}
& LUT & 301 & 196 & 34.9\% & 1.5x \\
 &  &  &  & FF  & 226 & 115 & 49.1\% & 2.0x \\
 &  &  &  & DSP & 1   & 0.5 & 50.0\% & 2.0x \\
\midrule
\multirow{3}{*}{FP4} & \multirow{3}{*}{FP16} & \multirow{3}{*}{FP16} & \multirow{3}{*}{FP16}
& LUT & 357 & 251 & 29.7\% & 1.4x \\
 &  &  &  & FF  & 266 & 131 & 50.8\% & 2.0x \\
 &  &  &  & DSP & 1   & 0.5 & 50.0\% & 2.0x \\
\midrule
\multirow{3}{*}{FP8} & \multirow{3}{*}{BF16} & \multirow{3}{*}{BF16} & \multirow{3}{*}{BF16}
& LUT & 301 & 219 & 27.2\% & 1.4x \\
 &  &  &  & FF  & 226 & 123 & 45.6\% & 1.8x \\
 &  &  &  & DSP & 1   & 0.5 & 50.0\% & 2.0x \\
\midrule
\multirow{3}{*}{FP8} & \multirow{3}{*}{FP16} & \multirow{3}{*}{FP16} & \multirow{3}{*}{FP16}
& LUT & 357 & 253 & 29.1\% & 1.4x \\
 &  &  &  & FF  & 266 & 133 & 50.0\% & 2.0x \\
 &  &  &  & DSP & 1   & 0.5 & 50.0\% & 2.0x \\
\bottomrule
\end{tabular}

\begin{tablenotes}[para,flushleft] \small
\item[$\dagger$] FP4 = E2M1, FP8 = E4M3.
\item[$\ddagger$] \name{} includes a non-blocking AXI wrapper for fair comparison. 
This introduces additional resource usage relative to Fig~\ref{fig:mixed_precision_matrix}, 
which reports only the core MAC datapath. 
\end{tablenotes}

\end{threeparttable}
}
\end{table}

\begin{figure}[t]
  \centering
  \setlength{\abovecaptionskip}{0.2cm}
  \setlength{\belowcaptionskip}{-0.2cm}
  \includegraphics[width=0.48\textwidth]{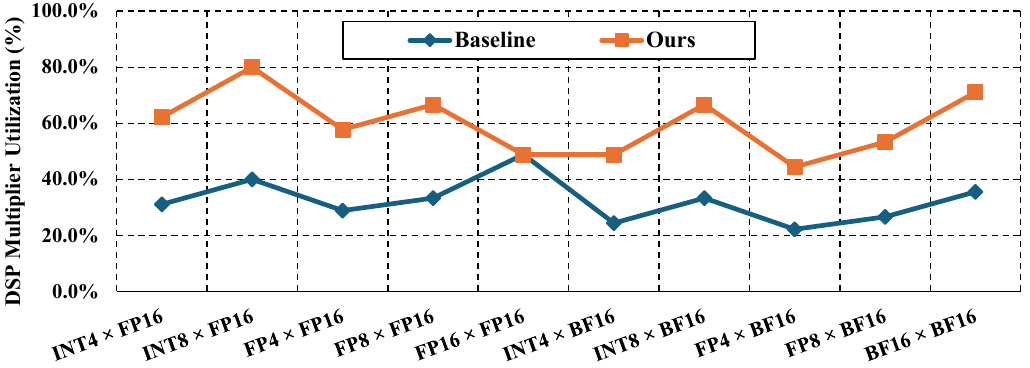}
  \caption{DSP utilization under different data types. (FP8=E4M3, FP4=E2M1).}
  \label{fig:dsp_usage}
  \vspace{-0.2cm}
\end{figure}

Table~\ref{tab:mixed_precision_comparison} reports resource consumption normalized by parallelism under representative mixed-precision operand formats; we divide each design’s total LUT, FF, and DSP usage by its number of MAC lanes to obtain per-operation utilization. Across these configurations, \name{} reduces LUT usage by an average of 30.0\%, FF usage by 47.9\%, and DSP usage by 50.0\% compared with the vendor IP. These reductions stem from \name’s adaptive parallelism for low-precision computation, which packs multiple mixed-precision lanes into a single DSP slice and increases effective multiplier utilization. Moreover, the bit-mapping datapath eliminates redundant bit operations and datatype-specific logic, reducing overall LUT and FF overhead. As illustrated in Fig.~\ref{fig:dsp_usage}, our data-packing strategy sustains high DSP utilization across all datatype combinations, enabling substantially more resource-efficient MAC execution than the vendor IP~\cite{floating_point_ip}. We further report compute density for LUTs, FFs, and DSPs as the ratio between the vendor IP resource usage and the \name{} usage; across all evaluated formats and resource types, \name{} achieves a compute density improvement between 1.4$\times$ and 2.0$\times$, indicating that each unit of hardware contributes more effective compute capability.

\begin{figure}[t]
  \centering
  \setlength{\abovecaptionskip}{0.2cm}
  \setlength{\belowcaptionskip}{-0.2cm}
  \includegraphics[width=0.48\textwidth]{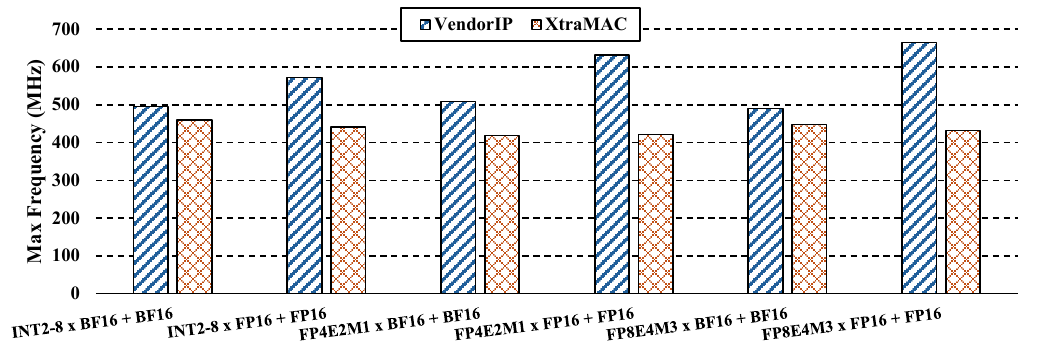}
  \caption{Maximum frequency comparison with a single DSP slice.}
  \label{fig:max_freq}
  \vspace{-0.2cm}
\end{figure}

We further evaluate the maximum frequency against the vendor IP with a single DSP slice. As shown in Fig.~\ref{fig:max_freq}, XtraMAC runs on average 22\% slower than the vendor IP~\cite{floating_point_ip} based MAC implementation. This reduction is expected: each XtraMAC configuration provides $2\times$ data parallelism per DSP compared to the $1\times$ baseline, which increases peripheral LUT and FF usage and introduces heavier routing congestion. Despite the frequency reduction, all XtraMAC configurations exceed 400\,MHz, well above the typical operating frequency of FPGA accelerator deployments. Because the $2\times$ parallelism doubles the per-DSP throughput, XtraMAC still delivers approximately $1.56\times$ higher effective throughput per DSP even after accounting for the frequency gap, confirming that the improved compute density more than compensates for the modest frequency overhead.

\subsubsection{Runtime datatype switching comparison}

We evaluate \name{}’s capability to switch between data formats dynamically during execution. Two baselines are included for comparison: the vendor’s IP~\cite{floating_point_ip} with spatial replication of INT8 and BF16 MAC units, and the TATAA accelerator~\cite{tataa2023}. All designs alternate between INT8-based and BF16-based MAC operations to emulate workloads that require runtime datatype switching. Resource per operation is obtained by dividing the total resources by the number of BF16 or INT8 lanes realized in each design.
For a fair comparison, all implementations use identical interface configurations and operate with an initiation interval (II) of~1 and a four-cycle latency.

\begin{table}[t]
\centering
\footnotesize
\caption{Normalized resource consumption per operation.}
\label{tab:normalized_resource}
\vspace{-0.2cm}
\resizebox{\linewidth}{!}{
\begin{threeparttable}
\begin{tabular}{
  >{\centering\arraybackslash}p{2.0cm} |
  >{\centering\arraybackslash}p{1.0cm}
  >{\centering\arraybackslash}p{1.0cm}
  >{\centering\arraybackslash}p{1.0cm} |
  >{\centering\arraybackslash}p{1.0cm}
  >{\centering\arraybackslash}p{1.0cm} 
  >{\centering\arraybackslash}p{1.0cm}
}
\toprule
\multirow{2}{*}{\textbf{Design}} &
\multicolumn{3}{c|}{\textbf{Resource per BF16 operation}} &
\multicolumn{3}{c}{\textbf{Resource per INT8 operation}} \\
\cmidrule(lr){2-4} \cmidrule(lr){5-7}
 & \textbf{LUT} & \textbf{FF} & \textbf{DSP}
 & \textbf{LUT} & \textbf{FF} & \textbf{DSP} \\
\midrule

Vendor IP~\cite{floating_point_ip}
& 220.0 & 310.5 & 1
& 110.0 & 155.3 & 0.5 \\
\midrule

TATAA~\cite{tataa2023}\tnote{$\dagger$}
& 352.0 & 467.0 & 4
& 22.0 & 29.2 & 0.25 \\
\midrule

\name
& 142.0 & 128.3 & 0.25
& 142.0 & 128.3 & 0.25 \\
\bottomrule
\end{tabular}

\begin{tablenotes}[flushleft]\small
\item[$\dagger$] For TATAA, we evaluate only the FP adder and multiplier and omit the \texttt{fapp} units (approximate sqrt/div), yielding lower resource usage than the full design in~\cite{tataa2023}.
\end{tablenotes}

\end{threeparttable}
}
\end{table}

Table~\ref{tab:normalized_resource} shows that \name{} provides substantial reductions in per-operation resource consumption under runtime precision switching. Relative to TATAA~\cite{tataa2023}, \name{} reduces LUT usage by 59.7\%, FF usage by 72.5\%, and DSP usage by 93.8\%. Relative to the vendor IP~\cite{floating_point_ip}, the reductions are 35.5\% for LUTs, 58.7\% for FFs, and 75.0\% for DSPs. These results indicate that \name{} achieves significantly higher hardware efficiency for floating-point computation while maintaining reasonable per-INT8 resource usage compared to INT8-optimized designs such as TATAA~\cite{tataa2023}. 

The inefficiency of the baselines in floating-point computation arises from how each design handles datatype heterogeneity. The vendor IP~\cite{floating_point_ip} spatially replicates independent INT8 and BF16 pipelines, preventing reuse of arithmetic or alignment logic across precisions and thereby duplicating hardware. TATAA~\cite{tataa2023} instead maps BF16 MACs onto its INT8 datapath, but must decompose each BF16 operation into multiple INT8 micro-operations with additional control and operand-adjustment hardware, which reduces utilization and BF16 efficiency. In contrast, \name{} employs a shared bit-mapping front-end that normalizes heterogeneous formats before entering the arithmetic core, so all precisions reuse the same multiplier–adder pipeline without duplication, substantially improving overall hardware efficiency under the same resource budget.

\section{Case Study: Mixed-Precision LLM Inference}
\label{sec:casestudy}

This section demonstrates how \name can be integrated into a practical tile-based GEMV pipeline and quantifies its benefits for mixed-precision LLM inference. We first describe the system-level integration of \name, then characterize the mixed-precision requirements of representative LLM workloads. Next, we evaluate a mixed-precision GEMV kernel based on \name and compare it against state-of-the-art GPU kernels. Finally, we extend an analytical simulation framework to estimate end-to-end LLM inference latency using \name-accelerated GEMV operators.

\subsection{Integration of \name into Tile-Based GEMV}

\begin{figure}[t]
  \centering
  \setlength{\abovecaptionskip}{0.2cm}
  \setlength{\belowcaptionskip}{-0.2cm}
  \includegraphics[width=0.48\textwidth]{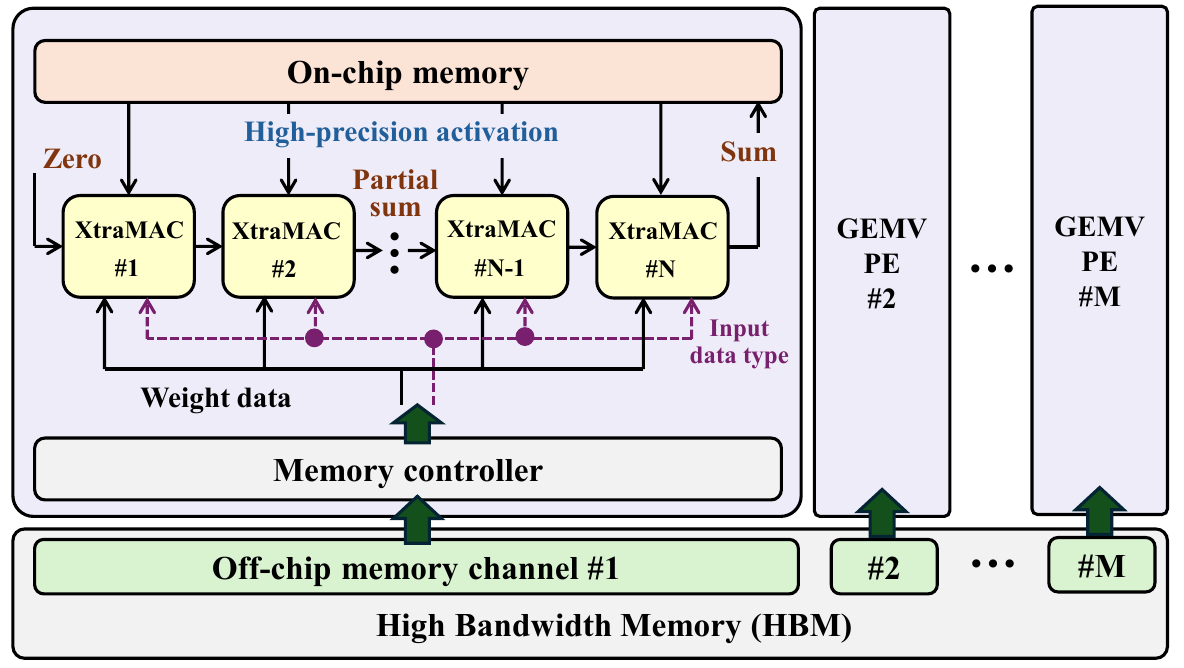}
  \caption{Mixed-precision GEMV architecture based on \name.}
  \label{fig:gemv}
  \vspace{-0.2cm}
\end{figure}

Figure~\ref{fig:gemv} illustrates how \name is incorporated into a tile-parallel GEMV engine. Modern FPGA GEMV pipelines~\cite{flightllm,flightvgm_fPGA25} implement streaming multiply--accumulate (MAC) chains within each processing element (PE), where weights are supplied directly from HBM and activations are buffered on-chip for reuse. \name preserves this interface and replaces the scalar MACs in prior designs without modifying the PE topology, tiling strategy, or global dataflow.

During execution, each wide memory-interface word (typically 256 or 512 bits) read from HBM is split into per-lane weight segments and dispatched to the corresponding \name{} instances within the PE. Each \name{} instance receives its weight segment alongside the corresponding activation value from the on-chip buffer. A per-tile datatype control signal, stored in memory alongside the weight tiles, is propagated synchronously with the operands to all \name{} units within the tile, ensuring that each instance applies the correct mapping rule and accumulation path at runtime. The resulting per-lane partial sums are accumulated across the cascaded MAC chain and written back as the final GEMV output, with multiple PEs operating concurrently across HBM channels to preserve bandwidth-scaling behavior.

\subsection{Mixed-Precision Requirements in LLM Workloads}

\begin{table}[t]
\centering
\footnotesize
\caption{Representative Quantized LLM Deployment Profiles.}
\label{tab:quantized_usecases_downloads}
\vspace{-0.3cm}
\begin{threeparttable}
\resizebox{0.98\linewidth}{!}{
\begin{tabular}{
  >{\centering\arraybackslash}m{1.8cm}|
  >{\centering\arraybackslash}m{1.8cm}|
  >{\raggedright\arraybackslash}m{6.0cm}
}
\toprule
\textbf{Checkpoints} & \textbf{Downloads*} & \textbf{Practical use cases}$^\dag$ \\
\midrule
Qwen-3-8B-AWQ & 222{,}126 & Edge AI reasoning with 2--3\% accuracy loss; long-context NLP with 91--97\% benchmark retention and a 4$\times$ context extension. \\
\midrule
Llama-3.1-8B-W8A8 & 27{,}536 & Commercial chatbots with 50\% memory reduction, 2$\times$ throughput, and 99--100\% accuracy for multilingual customer service. \\
\midrule
Qwen-3-8B-FP8 & 429{,}968 & Local agentic tasks and data analytics workflows; supports 131k-token contexts for scalable multilingual RAG. \\
\midrule
Llama-3.1-8B-FP8 & 168{,}122 & Multilingual assistants with 99.52\% accuracy recovery; optimized for fast commercial API deployments. \\
\midrule
GPT-oss-20B & 4{,}633{,}438 & Consumer fine-tuning, customizable inference latency, and low-latency analysis for agent-based data research. \\
\bottomrule
\end{tabular}
}
\begin{tablenotes}[para,flushleft]\footnotesize
\item[*] Download counts recorded on HuggingFace during October 2025. \\
\item[$\dag$] Accuracy loss in use cases only reflects model-level quantization effects.
\end{tablenotes}
\end{threeparttable}
\end{table}

Modern LLM deployments such as Llama~\cite{meta2024llama3}, Qwen~\cite{yang2025qwen3}, and GPT-oss~\cite{openai2025gptoss} heavily rely on quantization to reduce memory footprint and latency while preserving accuracy. Rather than adopting a single numeric format, practical systems combine integer and floating-point datatypes across layers. As summarized in Table~\ref{tab:patterns}, weight-only quantization uses integer weights with FP activations; weight--activation approaches quantize both operands; and mixed-format checkpoints (e.g., GPT-oss) interleave FP4/FP8 with BF16 for stability. Table~\ref{tab:quantized_usecases_downloads} profiles representative quantized checkpoints deployed in multilingual assistants, long-context RAG, and on-device LLMs. Collectively, these models indicate active deployment of INT4, FP4, and FP8 formats and highlight the need for GEMV engines that natively support mixed-precision and runtime datatype switching.

\subsection{Mixed-Precision GEMV Kernel Evaluation}

Across all evaluated models, these two patterns consistently dominate decode-time GEMV: (i) INT4$\times$BF16$\rightarrow$BF16 and (ii) FP4$\times$BF16$\rightarrow$BF16. These two patterns therefore serve as representative GEMV kernels for evaluating \name. All FPGA experiments are conducted on an AMD Xilinx U55c accelerator card equipped with 32 HBM channels. The \name{}-based GEMV kernel is synthesized and implemented using Vivado~2022.2 and Vitis~2022.2 with post-place-and-route timing. Power consumption is measured using \texttt{xbutil} under steady-state streaming. For comparison, GPU baselines are conducted on an NVIDIA H100 PCIe card using CUTLASS~\cite{cutlassProfile2025}; execution time is measured with official GEMV kernels, and power is monitored using \texttt{nvidia-smi}. This setup ensures a consistent and reproducible evaluation environment across both platforms.

To exploit HBM bandwidth, the GEMV workload is partitioned into $M$ tiles, each mapped to a processing element (PE) connected to a distinct HBM channel. Weights are stored in HBM, whereas activations are buffered on-chip. Inside each PE, the dot product is realized through a chain of cascaded \name{} modules. For a given HBM-channel bitwidth, weight precision, and datatype-dependent parallelism $P$, the number of cascaded \name{} instances per channel is determined by
\[
N_{\text{MAC}} = \frac{\text{BitWidth}_{\text{channel}}}{\text{BitWidth}_{\text{weight}} \times P}
\]
In implementation, each HBM channel provides a 512-bit interface; with INT4 weights (4\,bits) and $P=2$ lanes per \name{}, a single channel supplies $512/(4\times 2)=64$ MAC inputs per cycle. Thus, a design with 32 HBM channels can instantiate up to 2048 \name{} instances in principle, while our implementation adopts a configuration with 1920 cascaded XtraMAC instances (spread across 30 active HBM channels) to reserve one channel for activation read and one for write-back, and to ensure routing timing closure.

\begin{figure}[t]
  \centering
  \setlength{\abovecaptionskip}{0.2cm}
  \setlength{\belowcaptionskip}{-0.2cm}
  \includegraphics[width=0.48\textwidth]{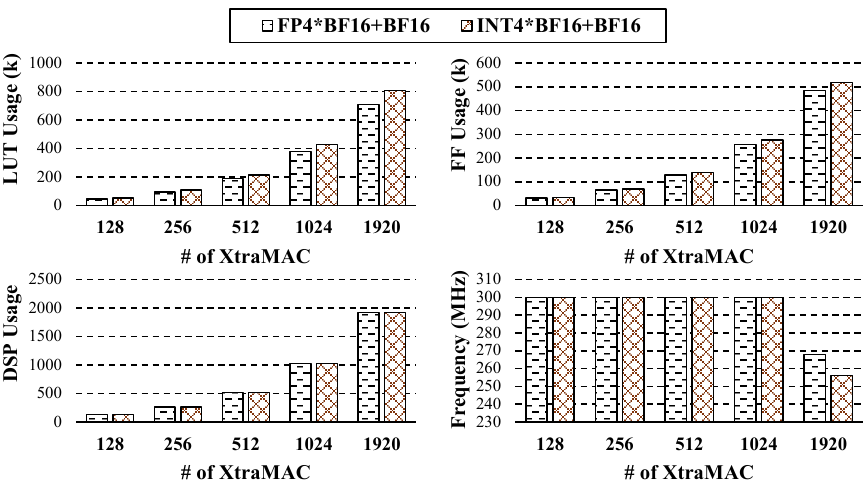}
  \caption{Post-implementation resource usage (LUT, FF, DSP) and frequency for mixed-precision GEMV kernel with various numbers of \name{}.}
  \label{fig:gemv_result}
  \vspace{-0.2cm}
\end{figure}

\begin{figure}[t]
  \centering
  \setlength{\abovecaptionskip}{0.3cm}
  \setlength{\belowcaptionskip}{-0.3cm}
  \includegraphics[width=0.45\textwidth]{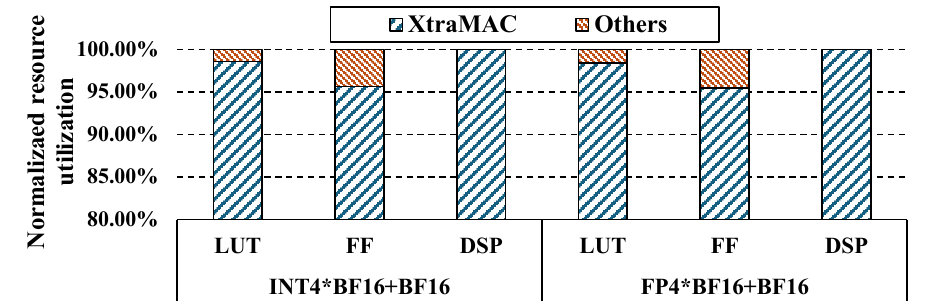}
  \caption{System-level resource breakdown (512-XtraMAC configuration).}
  \label{fig:breakdown}
  \vspace{-0.3cm}
\end{figure}

Post-implementation results in Figure~\ref{fig:gemv_result} indicate that LUT, FF, and DSP utilization scale linearly with the number of instantiated \name{} modules. The design sustains 300\,MHz across configurations up to 1024 instances and exhibits only moderate frequency degradation (250--270\,MHz) at 1920 instances due to routing congestion near the HBM interface. Because the kernel is bandwidth-bound at scale, these timing reductions have negligible impact on overall throughput. Figure~\ref{fig:breakdown} further presents the system-level resource breakdown under the 512-XtraMAC configuration. The XtraMAC instances dominate resource consumption, accounting for 98.5\% of LUTs, 95.6\% of FFs, and 100\% of DSPs. The remaining resources are consumed by supporting logic, including datatype control registers and on-chip activation buffers. HBM controllers reside in the Alveo static shell~\cite{amd_u55c_platform} and are excluded from user-logic resource accounting. All non-MAC components are registered in parallel with the operand datastream and introduce no additional pipeline stages, preserving the fixed latency and II\,=\,1 of the MAC pipeline.

Table~\ref{tab:mixed_gemv_perf} presents the performance comparison against the H100 GPU. Despite the GPU’s substantially higher peak memory bandwidth (2\,TB/s versus 460\,GB/s), the \name{}-based FPGA kernel achieves 1.2$\times$ lower latency and 1.9$\times$ higher energy efficiency on the $1\times 4096 \times 4096$ and $1\times 4096 \times 12288$ GEMV workloads. These gains arise from the efficient mixed-precision MAC datapath in \name, the absence of format-conversion overheads, and the kernel’s ability to sustain high effective HBM utilization (approximately 74\%). As a result, the FPGA design operates close to its bandwidth roofline and surpasses the GPU on bandwidth-bound GEMV workloads.

\begin{table}[t]
\centering
\footnotesize
\caption{Comparison of mixed-precision GEMV performance.}
\vspace{-0.2cm}
\label{tab:mixed_gemv_perf}
\resizebox{\linewidth}{!}{
\begin{tabular}{l|c|c|c|c|c}
\toprule
\multicolumn{6}{c}{\textbf{$1\times4096\times4096$ GEMV}} \\
\midrule
\textbf{Design} & \textbf{Time (ms)} & \textbf{Power (W)} & \textbf{Energy (J)} & \textbf{Speedup} & \textbf{Energy Eff.} \\
\midrule
CUTLASS (H100)            & 0.0294 & 135 & 0.00397  & 1.0$\times$ & 1.0$\times$ \\
\textbf{\name{} (U55c)}   & \textbf{0.0246} & \textbf{85}  & \textbf{0.00209} & \textbf{1.2$\times$} & \textbf{1.9$\times$} \\
\midrule
\multicolumn{6}{c}{\textbf{$1\times4096\times12288$ GEMV}} \\
\midrule
\textbf{Design} & \textbf{Time (ms)} & \textbf{Power (W)} & \textbf{Energy (J)} & \textbf{Speedup} & \textbf{Energy Eff.} \\
\midrule
CUTLASS (H100)            & 0.0879 & 135 & 0.01187  & 1.0$\times$ & 1.0$\times$ \\
\textbf{\name{} (U55c)}   & \textbf{0.0743} & \textbf{85}  & \textbf{0.00632} & \textbf{1.2$\times$} & \textbf{1.9$\times$} \\
\bottomrule
\end{tabular}}
\end{table}

\subsection{End-to-End Mixed-Precision LLM Inference Simulation}

\begin{figure}[t]
  \centering
  \setlength{\abovecaptionskip}{0.2cm}
  \setlength{\belowcaptionskip}{-0.2cm}
  \includegraphics[width=0.47\textwidth]{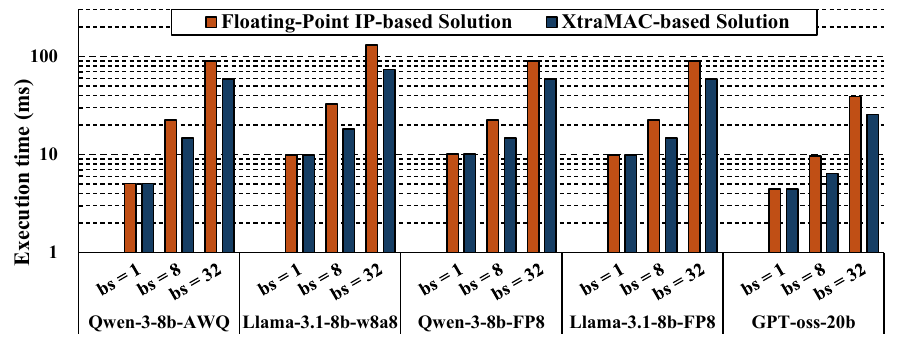}
  \caption{Decode-stage execution time for different LLMs at context length 512 and batch sizes $\{1,8,32\}$.}
  \label{fig:llm_inference}
  \vspace{-0.2cm}
\end{figure}

We evaluate system-level behavior using the analytical framework of~\cite{spatial_llm_fpga}, which models transformer layers as alternating compute and memory phases under idealized streaming and weight-reuse assumptions. All models in Table~\ref{tab:quantized_usecases_downloads} are simulated on the AMD Alveo V80 accelerator (2.6M LUTs, 10{,}848 DSPs, 300\,MHz, and 810\,GB/s HBM). The baseline instantiates the vendor Floating-Point IP using the resource profiles in Table~\ref{tab:mixed_precision_comparison}, while our configuration replaces these units with \name. All other components, including datatype packing rules, checkpoint-derived MAC counts, and tiling behavior, are kept fixed so that performance differences reflect only the arithmetic-unit density.

Figure~\ref{fig:llm_inference} reports decode latency at a context length of 512 for batch sizes $\{1, 8, 32\}$. At batch size~1, all five evaluated models (Qwen-3-8B-AWQ, Llama-3.1-8B-W8A8, Qwen-3-8B-FP8, Llama-3.1-8B-FP8, and GPT-oss-20B) operate in a memory-bound regime. Weight streaming dominates the end-to-end latency (approximately 4.4--10.0\,ms), and compute differences between the baseline and \name{} have negligible impact. As the batch size increases, the compute workload grows while the weight-fetch cost remains nearly constant, placing all models in a compute-bound regime at batch sizes~8 and~32. In this region, the reduced LUT and DSP footprint of \name{} enables more MAC units to be instantiated on the FPGA platform, which produces consistent reductions in total latency. At batch size~32, \name{} delivers improvements ranging from 1.5$\times$ to 1.8$\times$ for the end-to-end LLM inference task. These trends align with the proportion of mixed-precision GEMV operations in each model and confirm that arithmetic-unit density, rather than memory bandwidth, becomes the dominant performance limiter at large batch sizes.

\subsection{Design Guidelines} 

In this subsection, we summarize several guidelines for integrating \name{} into system-level designs.

First, \name{} maintains a constant latency and an initiation interval of one across all supported datatypes, and can therefore serve as a drop-in replacement for scalar MAC units in existing GEMV/GEMM pipelines, requiring no changes to pipeline depth, scheduling logic, or interface timing. Second, runtime datatype switching is equally straightforward: a per-tile control signal propagated synchronously with the operand datastream is sufficient, so mixed-precision workloads can be scheduled as if all datatypes share a single unified pipeline, without the need for pipeline flushes or reconfiguration. Last, from a quantization perspective, lane packing efficiency scales inversely with operand bitwidth. Sub-8-bit quantization schemes (e.g., INT4, FP4) can therefore be prioritized in FPGA deployments to maximize per-DSP parallelism, directly translating into higher overall throughput.

\section{Related Work}

In this section, we review existing approaches to mixed-precision MAC computation and runtime datatype switching, and position \name{} against their remaining limitations.

\subsection{Fixed-Precision Designs}
General-purpose processors expose multiple numerical formats via fixed hardware pipelines. Modern CPUs support integer and low-precision floating-point arithmetic through instruction-set extensions such as AVX-512, AMX, and ARM SVE, while GPUs accelerate machine learning workloads using Tensor Cores with support for FP16, BF16, FP8, and INT8~\cite{nvidia_a100_ds, nvidia2022h100}. Domain-specific accelerators such as Google TPU~\cite{jouppi2017tpu, tpu_v5e, tpu_ironwood} employ systolic arrays optimized for a fixed set of precisions (e.g., BF16 and INT8), with each datatype served by dedicated compute pipelines. However, across all these designs, mixed-precision computation is handled either by upcasting low-precision operands to a common high-precision format, or by routing different datatypes through physically separated execution pipelines. Both strategies leave hardware resources idle when only a subset of datatypes is active, and neither exploits the fine-grained bit-level parallelism available within a shared datapath.

\subsection{Configurable-Precision Designs}

A second class of designs introduces runtime precision adaptivity. BitFusion~\cite{sharma2018bitfusion} fuses 2-bit processing elements dynamically to match the integer bitwidth of each DNN layer. Stripes~\cite{judd2016stripes} employs bit-serial computation so that execution latency scales proportionally with operand precision. OLAccel~\cite{ma2022olaccel} supports operand-level precision variability through heterogeneous MAC units and temporal decomposition. 
More recently, FlexiBit~\cite{flexibit2024} proposes a configurable bit-parallel reduction tree supporting arbitrary FP and INT precisions, including non-power-of-two formats. However, BitFusion~\cite{sharma2018bitfusion}, Stripes~\cite{judd2016stripes}, and OLAccel~\cite{ma2022olaccel} support only integer formats and lack native floating-point computation. FlexiBit, while supporting FP formats, is an ASIC design and thus requires hardware re-fabrication to accommodate new quantization schemes. None of these designs support cycle-level runtime switching between heterogeneous INT and FP formats within a shared datapath.

\name{} bridges these gaps by unifying integer, floating-point, and mixed-precision multiplication within a single resource-compact FPGA datapath. By reducing all MAC formats to a shared integer mantissa product and decoupling accumulation paths, \name{} achieves cycle-level runtime datatype switching with constant latency and high throughput.

\section{Conclusion}
MAC operations dominate the computational cost of modern LLM inference, yet existing FPGA MAC designs remain inefficient because they either upcast low-precision operands or replicate datatype-specific datapaths, resulting in poor DSP utilization and limited runtime flexibility. This work introduces \name, a compact, datatype-adaptive MAC architecture that supports mixed-precision computation with cycle-level runtime switching. The key architectural novelty of \name lies in unifying INT and FP formats through a shared integer-product formulation, enabling multiple low-precision lanes to be packed within a single DSP while maintaining a fixed latency. Compared with state-of-the-art MAC designs, \name significantly improves hardware efficiency, delivering 1.4--2.0$\times$ higher compute density while reducing LUT, FF, and DSP usage by 30.0\%, 47.9\%, and 50.0\%, respectively. 

\section*{Acknowledgment}
This work is supported by the Ministry of Education AcRF Tier 3 grant, Singapore (MOE-MOET32024-0003), and a Google Gift 2025. We also thank the AMD Heterogeneous Accelerated Compute Clusters (HACC) program \cite{hacc} for the generous hardware donation.

\bibliographystyle{IEEEtranS}
\bibliography{refs}

@misc{nvidia_a100_ds,
  title        = {{NVIDIA} {A100} Tensor Core {GPU} Datasheet},
  author       = {{NVIDIA Corporation}},
  howpublished = {\url{https://www.nvidia.com/content/dam/en-zz/Solutions/Data-Center/a100/pdf/nvidia-a100-datasheet-us-nvidia-1758950-r4-web.pdf}},
  year         = {2021},
  note         = {Accessed Oct. 2025}
}

@misc{tpu_v5e,
  title        = {Cloud {TPU} v5e},
  author       = {{Google Cloud}},
  howpublished = {\url{https://cloud.google.com/tpu/docs/v5e}},
  note         = {Accessed Oct. 2025}
}

@misc{tpu_ironwood,
  title        = {Ironwood: The First {Google} {TPU} for the Age of Inference},
  author       = {Vahdat, Amin},
  howpublished = {\url{https://blog.google/products/google-cloud/ironwood-tpu-age-of-inference/}},
  note         = {Accessed Oct. 2025}
}

@manual{floating_point_ip,
  title        = {{AMD} {Xilinx} Floating-Point Operator v7.1 {LogiCORE} {IP} Product Guide ({PG060})},
  organization = {{Advanced Micro Devices, Inc.}},
  url          = {https://docs.amd.com/v/u/en-US/pg060-floating-point},
  note         = {Accessed Oct. 2025}
}

@misc{amd_u55c_platform,
  title        = {{Alveo} {U55C} Accelerator Card: Support and Downloads},
  author       = {{AMD}},
  howpublished = {\url{https://www.amd.com/en/products/accelerators/alveo/u55c/a-u55c-p00g-pq-g.html}},
  note         = {Accessed Oct. 2025}
}

@article{arora2022tensor,
  author    = {Arora, Aman and Ghosh, Moinak and Mehta, Samidh and Betz, Vaughn and John, Lizy K.},
  title     = {Tensor Slices: {FPGA} Building Blocks for the Deep Learning Era},
  journal   = {{ACM} Trans. Reconfigurable Technol. Syst.},
  volume    = {15},
  number    = {4},
  articleno = {46},
  numpages  = {34},
  year      = {2022},
  month     = Dec,
  doi       = {10.1145/3529650}
}

@inproceedings{chee2023quip,
author = {Chee, Jerry and Cai, Yaohui and Kuleshov, Volodymyr and De Sa, Christopher},
title = {QuIP: 2-bit quantization of large language models with guarantees},
year = {2023},
publisher = {Curran Associates Inc.},
address = {Red Hook, NY, USA},
booktitle = {Proceedings of the 37th International Conference on Neural Information Processing Systems},
articleno = {196},
numpages = {34},
location = {New Orleans, LA, USA},
series = {NIPS '23}
}

@article{spatial_llm_fpga,
author = {Chen, Hongzheng and Zhang, Jiahao and Du, Yixiao and Xiang, Shaojie and Yue, Zichao and Zhang, Niansong and Cai, Yaohui and Zhang, Zhiru},
title = {Understanding the Potential of FPGA-based Spatial Acceleration for Large Language Model Inference},
year = {2024},
issue_date = {March 2025},
publisher = {Association for Computing Machinery},
address = {New York, NY, USA},
volume = {18},
number = {1},
issn = {1936-7406},
doi = {10.1145/3656177},
journal = {ACM Trans. Reconfigurable Technol. Syst.},
month = dec,
articleno = {5},
numpages = {29}
}

@inproceedings{hipack,
  title     = {{HiPACK}: Efficient Sub-8-Bit Direct Convolution with {SIMD} and Bitwise Management},
  author    = {Chen, Yao and Gong, Cheng and He, Bingsheng},
  booktitle = {Proceedings of the 58th {IEEE/ACM} International Symposium on Microarchitecture ({MICRO})},
  pages     = {1579--1591},
  year      = {2025},
  doi       = {10.1145/3725843.3756124}
}

@INPROCEEDINGS{chen2023m4bram,
  author={Chen, Yuzong and Dotzel, Jordan and Abdelfattah, Mohamed S.},
  booktitle={2023 International Conference on Field Programmable Technology (ICFPT)}, 
  title={M4BRAM: Mixed-Precision Matrix-Matrix Multiplication in FPGA Block RAMs}, 
  year={2023},
  volume={},
  number={},
  pages={69-78},
  doi={10.1109/ICFPT59805.2023.00013}}

@misc{int_to_float,
  title        = {Integer-to-Floating Point Converter ({Verilog} Implementation)},
  author       = {Dawson, Jonathan},
  howpublished = {\url{https://github.com/dawsonjon/fpu/blob/master/int_to_float/int_to_float.v}},
  note         = {Accessed Oct. 2025}
}

@article{dettmers2024spqr,
  title={Spqr: A sparse-quantized representation for near-lossless llm weight compression},
  author={Dettmers, Tim and Svirschevski, Ruslan and Egiazarian, Vage and Kuznedelev, Denis and Frantar, Elias and Ashkboos, Saleh and Borzunov, Alexander and Hoefler, Torsten and Alistarh, Dan},
  journal={arXiv preprint arXiv:2306.03078},
  year={2023}
}

@article{meta2024llama3,
  title   = {The {Llama 3} Herd of Models},
  author  = {Grattafiori, Aaron and Dubey, Abhimanyu and others},
  journal = {arXiv preprint arXiv:2407.21783},
  year    = {2024},
  url     = {https://arxiv.org/abs/2407.21783}
}

@inproceedings{ehliar2014area,
  title     = {Area Efficient Floating-Point Adder and Multiplier with {IEEE-754} Compatible Semantics},
  author    = {Ehliar, Andreas},
  booktitle = {2014 International Conference on Field-Programmable Technology ({FPT})},
  pages     = {131--138},
  year      = {2014},
  publisher = {IEEE},
  doi       = {10.1109/FPT.2014.7082765}
}

@article{frantar2022gptq,
  title   = {{GPTQ}: Accurate Post-Training Quantization for Generative Pretrained Transformers},
  author  = {Frantar, Elias and Ashkboos, Saleh and Hoefler, Torsten and Alistarh, Dan},
  journal = {arXiv preprint arXiv:2210.17323},
  year    = {2022},
  url     = {https://arxiv.org/abs/2210.17323}
}

@techreport{xilinx_int8_wp486,
  title       = {Deep Learning with {INT8} Optimization on {Xilinx} Devices},
  author      = {Fu, Yao and Wu, Ephrem and Sirasao, Ashish and Attia, Sedny and Khan, Kamran and Wittig, Ralph},
  institution = {{Xilinx, Inc.}},
  type        = {White Paper},
  number      = {WP486 v1.0.1},
  year        = {2017},
  month       = {April},
  url         = {https://docs.amd.com/api/khub/documents/z7yAy_aweTmRYkGaTVyhbw/content}
}

@inproceedings{bismo2018,
  author    = {Umuroglu, Yaman and Rasnayake, Lahiru and Sj{\"a}lander, Magnus},
  title     = {{BISMO}: A Scalable Bit-Serial Matrix Multiplication Overlay for Reconfigurable Computing},
  booktitle = {2018 28th International Conference on Field Programmable Logic and Applications ({FPL})},
  pages     = {307--314},
  year      = {2018},
  publisher = {IEEE},
  doi       = {10.1109/FPL.2018.00059}
}

@mastersthesis{hartman1996mmachine,
  author  = {Hartman, Daniel K.},
  title   = {Floating Point Multiply/Add Unit for the {M-Machine} Node Processor},
  school  = {Massachusetts Institute of Technology},
  address = {Cambridge, MA},
  year    = {1996},
  month   = {May},
  url     = {http://hdl.handle.net/1721.1/38791}
}

@inproceedings{jouppi2017tpu,
  title={In-datacenter performance analysis of a tensor processing unit},
  author={Jouppi, Norman P and Young, Cliff and Patil, Nishant and Patterson, David and Agrawal, Gaurav and Bajwa, Raminder and Bates, Sarah and Bhatia, Suresh and Boden, Nan and Borchers, Al and others},
  booktitle={Proceedings of the 44th annual international symposium on computer architecture (ISCA)},
  pages={1--12},
  year={2017}
}

@inproceedings{judd2016stripes,
  title     = {{Stripes}: Bit-Serial Deep Neural Network Computing},
  author    = {Judd, Patrick and Albericio, Jorge and Hetherington, Tayler H. and Aamodt, Tor M. and Moshovos, Andreas},
  booktitle = {Proceedings of the 49th Annual {IEEE/ACM} International Symposium on Microarchitecture ({MICRO})},
  pages     = {19:1--19:12},
  year      = {2016},
  doi       = {10.1109/MICRO.2016.7783722}
}

@inproceedings{kerner2021triple,
  title     = {Triple Fixed-Point {MAC} Unit for Deep Learning},
  author    = {Kerner, Madis and Tammem{\"a}e, Kalle and Raik, Jaan and Hollstein, Thomas},
  booktitle = {2021 Design, Automation \& Test in Europe Conference \& Exhibition ({DATE})},
  pages     = {1404--1407},
  year      = {2021},
  doi       = {10.23919/DATE51398.2021.9474020}
}

@article{liang2017fpbnn,
  title   = {{FP-BNN}: Binarized Neural Network on {FPGA}},
  author  = {Liang, Shuang and Yin, Shouyi and Liu, Leibo and Luk, Wayne and Wei, Shaojun},
  journal = {Neurocomputing},
  volume  = {275},
  pages   = {1072--1086},
  year    = {2018},
  doi     = {10.1016/j.neucom.2017.09.046}
}

@inproceedings{lin2024awq,
  author    = {Lin, Ji and Tang, Jiaming and Tang, Haotian and Yang, Shang and Chen, Wei-Ming and Wang, Wei-Chen and Xiao, Guangxuan and Dang, Xingyu and Gan, Chuang and Han, Song},
  title     = {{AWQ}: Activation-Aware Weight Quantization for On-Device {LLM} Compression and Acceleration},
  booktitle = {Proceedings of Machine Learning and Systems ({MLSys})},
  pages     = {87--100},
  volume    = {6},
  year      = {2024}
}

@inproceedings{flightvgm_fPGA25,
  author    = {Liu, Jun and Zeng, Shulin and Ding, Li and Soedarmadji, Widyadewi and Zhou, Hao and Wang, Zehao and Li, Jinhao and Li, Jintao and Dai, Yadong and Wen, Kairui and He, Shan and Sun, Yaqi and Wang, Yu and Dai, Guohao},
  title     = {{FlightVGM}: Efficient Video Generation Model Inference with Online Sparsification and Hybrid Precision on {FPGAs}},
  booktitle = {Proceedings of the {ACM/SIGDA} International Symposium on Field Programmable Gate Arrays ({FPGA})},
  pages     = {2--13},
  year      = {2025},
  doi       = {10.1145/3706628.3708864}
}

@inproceedings{Hikonv,
  author    = {Liu, Xinheng and Chen, Yao and Ganesh, Prakhar and Pan, Junhao and Xiong, Jinjun and Chen, Deming},
  title     = {{HiKonv}: High Throughput Quantized Convolution with Novel Bit-Wise Management and Computation},
  booktitle = {2022 27th Asia and South Pacific Design Automation Conference ({ASP-DAC})},
  pages     = {140--146},
  year      = {2022},
  doi       = {10.1109/ASP-DAC52403.2022.9712553}
}

@inproceedings{ma2022olaccel,
  title     = {Energy-Efficient Neural Network Accelerator Based on Outlier-Aware Low-Precision Computation},
  author    = {Park, Eunhyeok and Kim, Dongyoung and Yoo, Sungjoo},
  booktitle = {Proceedings of the 45th Annual International Symposium on Computer Architecture ({ISCA})},
  pages     = {688--698},
  year      = {2018},
  doi       = {10.1109/ISCA.2018.00063}
}

@inproceedings{moctar2012reducing,
  title     = {Reducing the Cost of Floating-Point Mantissa Alignment and Normalization in {FPGAs}},
  author    = {Moctar, Yehdhih Ould Mohammed and George, Nithin and Parandeh-Afshar, Hadi and Ienne, Paolo and Lemieux, Guy GF and Brisk, Philip},
  booktitle = {Proceedings of the {ACM/SIGDA} International Symposium on Field-Programmable Gate Arrays ({FPGA})},
  pages     = {255--264},
  year      = {2012},
  doi       = {10.1145/2145694.2145738}
}

@techreport{nvidia_ampere_whitepaper,
  title       = {{NVIDIA} {A100} Tensor Core {GPU} Architecture},
  author      = {{NVIDIA Corporation}},
  institution = {{NVIDIA}},
  year        = {2020},
  month       = {May},
  type        = {White Paper},
  url         = {https://resources.nvidia.com/en-us-tensor-core/nvidia-ampere-architecture-whitepaper}
}

@techreport{nvidia2022h100,
  title       = {{NVIDIA} {H100} Tensor Core {GPU} Architecture},
  author      = {{NVIDIA Corporation}},
  institution = {{NVIDIA}},
  year        = {2022},
  type        = {White Paper},
  url         = {https://resources.nvidia.com/en-us-tensor-core/nvidia-h100-whitepaper}
}

@misc{cutlassProfile2025,
  title        = {{CUTLASS} 3.x Performance Profiling Results},
  author       = {{NVIDIA Corporation}},
  howpublished = {\url{https://github.com/NVIDIA/cutlass}},
  year         = {2025},
  note         = {Accessed Oct. 2025}
}

@article{openai2025gptoss,
  title   = {{gpt-oss-120b} \& {gpt-oss-20b} Model Card},
  author  = {{OpenAI}},
  journal = {arXiv preprint arXiv:2508.10925},
  year    = {2025},
  url     = {https://arxiv.org/abs/2508.10925}
}

@inproceedings{schulte2005barrel,
  title={Design alternatives for barrel shifters},
  author={Pillmeier, Matthew R and Schulte, Michael J and Walters III, Eugene George},
  booktitle={Advanced Signal Processing Algorithms, Architectures, and Implementations XII},
  volume={4791},
  pages={436--447},
  year={2002},
  organization={SPIE}
}

@inproceedings{sharma2018bitfusion,
  title     = {{Bit Fusion}: Bit-Level Dynamically Composable Architecture for Accelerating Deep Neural Network},
  author    = {Sharma, Hardik and Park, Jongse and Suda, Naveen and Lai, Liangzhen and Chau, Benson and Chandra, Vikas and Esmaeilzadeh, Hadi},
  booktitle = {Proceedings of the 45th Annual International Symposium on Computer Architecture ({ISCA})},
  pages     = {764--775},
  year      = {2018},
  publisher = {IEEE},
  doi       = {10.1109/ISCA.2018.00069}
}

@misc{flexibit2024,
  author        = {Tahmasebi, Faraz and Wang, Yian and Huang, Benji Y. H. and Kwon, Hyoukjun},
  title         = {{FlexiBit}: Fully Flexible Precision Bit-Parallel Accelerator Architecture for Arbitrary Mixed Precision {AI}},
  year          = {2024},
  eprint        = {2411.18065},
  archivePrefix = {arXiv},
  primaryClass  = {cs.AR},
  url           = {https://arxiv.org/abs/2411.18065}
}

@inproceedings{umuroglu2017finn,
  title     = {{FINN}: A Framework for Fast, Scalable Binarized Neural Network Inference},
  author    = {Umuroglu, Yaman and Fraser, Nicholas J. and Gambardella, Giulio and Blott, Michaela and Leong, Philip and Jahre, Magnus and Vissers, Kees},
  booktitle = {Proceedings of the {ACM/SIGDA} International Symposium on Field-Programmable Gate Arrays ({FPGA})},
  pages     = {65--74},
  year      = {2017},
  doi       = {10.1145/3020078.3021744}
}

@article{tataa2023,
  author    = {Wu, Jiajun and Song, Mo and Zhao, Jingmin and Gao, Yizhao and Li, Jia and So, Hayden Kwok-Hay},
  title     = {{TATAA}: Programmable Mixed-Precision Transformer Acceleration with a Transformable Arithmetic Architecture},
  journal   = {{ACM} Trans. Reconfigurable Technol. Syst.},
  volume    = {18},
  number    = {1},
  articleno = {14},
  pages     = {14:1--14:31},
  year      = {2025},
  doi       = {10.1145/3714416}
}

@inproceedings{smoothquant,
  title={Smoothquant: Accurate and efficient post-training quantization for large language models},
  author={Xiao, Guangxuan and Lin, Ji and Seznec, Mickael and Wu, Hao and Demouth, Julien and Han, Song},
  booktitle={International conference on machine learning},
  pages={38087--38099},
  year={2023},
  organization={PMLR}
}

@misc{xilinx_dsp48e2_ug958,
  title       = {Vivado Design Suite Reference Guide: Model-Based DSP Design Using System Generator (UG958)},
  author      = {{AMD Xilinx, Inc.}},
  institution = {{AMD Xilinx, Inc.}},
  url         = {https://docs.amd.com/r/en-US/ug958-vivado-sysgen-ref/DSP48E2},
  note         = {Accessed Oct. 2025}
}

@misc{hacc,
  title        = {Heterogeneous Accelerated Compute Cluster ({HACC}) at {NUS}},
  author       = {{AMD Xilinx}},
  howpublished = {\url{https://xacchead.ddns.comp.nus.edu.sg/}},
  note         = {Accessed Oct. 2025}
}

@manual{ug579,
  title        = {{UltraScale} Architecture {DSP} Slice User Guide ({UG579})},
  author       = {{Xilinx, Inc.}},
  organization = {{Xilinx, Inc.}},
  year         = {2023},
  url          = {https://docs.xilinx.com/v/u/en-US/ug579-ultrascale-dsp}
}

@article{yang2025qwen3,
  title   = {{Qwen3} Technical Report},
  author  = {Yang, An and others},
  journal = {arXiv preprint arXiv:2505.09388},
  year    = {2025},
  url     = {https://arxiv.org/abs/2505.09388}
}

@inproceedings{flightllm,
  title     = {{FlightLLM}: Efficient Large Language Model Inference with a Complete Mapping Flow on {FPGAs}},
  author    = {Zeng, Shulin and Liu, Jun and Dai, Guohao and Yang, Xinhao and Fu, Tianyu and Wang, Hongyi and Ma, Wenheng and Sun, Hanbo and Li, Shiyao and Huang, Zixiao and Dai, Yadong and Li, Jintao and Wang, Zehao and Zhang, Ruoyu and Wen, Kairui and Ning, Xuefei and Wang, Yu},
  booktitle = {Proceedings of the {ACM/SIGDA} International Symposium on Field Programmable Gate Arrays ({FPGA})},
  pages     = {223--234},
  year      = {2024},
  doi       = {10.1145/3626202.3637562}
}

@inproceedings{zhang2018dnnbuilder,
  title     = {{DNNBuilder}: An Automated Tool for Building High-Performance {DNN} Hardware Accelerators for {FPGAs}},
  author    = {Zhang, Xiaofan and Wang, Junsong and Zhu, Chao and Lin, Yonghua and Xiong, Jinjun and Hwu, Wen-mei and Chen, Deming},
  booktitle = {Proceedings of the {IEEE/ACM} International Conference on Computer-Aided Design ({ICCAD})},
  articleno = {56},
  pages     = {1--8},
  year      = {2018},
  doi       = {10.1145/3240765.3240801}
}

@inproceedings{MLSYS2024_atom,
  author    = {Zhao, Yilong and Lin, Chien-Yu and Zhu, Kan and Ye, Zihao and Chen, Lequn and Zheng, Size and Ceze, Luis and Krishnamurthy, Arvind and Chen, Tianqi and Kasikci, Baris},
  title     = {{Atom}: Low-Bit Quantization for Efficient and Accurate {LLM} Serving},
  booktitle = {Proceedings of Machine Learning and Systems ({MLSys})},
  pages     = {196--209},
  volume    = {6},
  year      = {2024}
}

\end{document}